\documentclass{JHEP3}
\usepackage{bm}
\usepackage{amsmath}
\usepackage{epsfig}

\title{Can residuals of the Solar system foreground explain low multipole anomalies of the CMB ?}

\author{M. Hansen$^1$, J. Kim$^1$, A.M. Frejsel$^1$, S.Ramazanov$^{1,2}$ P. Naselsky$^1$, W. Zhao$^1$ and C. Burigana$^{3,4}$ \\
$^1$ Niels Bohr Institute and DISCOVERY Center, Blegdamsvej 17, 2100 Copenhagen, {\O}, Denmark\\
$^2$ Moscow State University, 11999, Moscow, Russia\\
$^3$ INAF/IASF, Istituto di Astrofisica Spaziale e Fisica Cosmica di Bologna,Via P. Gobetti 101, 40129, Bologna, Italy\\
$^4$ Dipartimento di Fisica, Universit\`a degli Studi di Ferrara, Via G. Saragat 1, 44100 Ferrara, Italy}

\abstract{
The low multipole anomalies of the Cosmic Microwave Background has received much attention during the last few years. It is still not ascertained whether these anomalies are indeed primordial or the result of systematics or foregrounds. An example of a foreground, which could generate some non-Gaussian and statistically anisotropic features at low multipole range, is the very symmetric Kuiper Belt in the outer solar system. In this paper, expanding upon the methods presented in \cite{Burigana}, we investigate the contributions from the Kuiper Belt objects (KBO) to the WMAP ILC 7 map, whereby we can minimize the contrast in power between even and odd multipoles in the CMB, discussed in \cite{Jkim1,Jkim2}.\\
We submit our KBO de-correlated CMB signal to several tests, to analyze its validity, and find that incorporation of the KBO emission can decrease the quadrupole-octupole alignment and parity asymmetry problems, provided that the KBO signals has a non-cosmological dipole modulation, associated with the statistical anisotropy of the ILC 7 map.  Additionally, we show that the amplitude of the dipole modulation, within a $2\sigma$ interval, is in agreement with the corresponding amplitudes, discussed in \cite{Lew}.
}

\makeindex
\preprint{-}
\begin{document}

\newcommand{\apjs}{ApJS}
\newcommand{\mnras}{MNRAS}

\newcommand{\beq}{\begin{equation}}
\newcommand{\eeq}{\end{equation}}
\newcommand{\be}{\begin{eqnarray}}
\newcommand{\ee}{\end{eqnarray}}
\newcommand{\num}{\nu_\mu}
\newcommand{\nue}{\nu_e}
\newcommand{\nut}{\nu_\tau}
\newcommand{\nus}{\nu_s}
\newcommand{\mnus}{M_s}
\newcommand{\taus}{\tau_{\nu_s}}
\newcommand{\nnt}{n_{\nu_\tau}}
\newcommand{\rnt}{\rho_{\nu_\tau}}
\newcommand{\mnt}{m_{\nu_\tau}}
\newcommand{\tnt}{\tau_{\nu_\tau}}
\newcommand{\rar}{\rightarrow}
\newcommand{\lar}{\leftarrow}
\newcommand{\lrar}{\leftrightarrow}
\newcommand{\dm}{\delta m^2}
\newcommand{\mpl}{m_{Pl}}
\newcommand{\mbh}{M_{BH}}
\newcommand{\nbh}{n_{BH}}
\newcommand{\crit}{{\rm crit}}
\newcommand{\ini}{{\rm in}}
\newcommand{\cmb}{{\rm cmb}}
\newcommand{\rec}{{\rm rec}}
\newcommand{\lsim} {\mathrel{\hbox{\rlap{\lower.55ex \hbox{$\sim$}}
                             \kern-.3em \raise.4ex \hbox{$<$}}}} 

\newcommand{\Odm}{\Omega_{\rm dm}}
\newcommand{\Ob}{\Omega_{\rm b}}
\newcommand{\Om}{\Omega_{\rm m}}
\newcommand{\nb}{n_{\rm b}}
\def\simlt{\lesssim}
\def\simgt{\gtrsim}
\def\Cl{C_{\ell}}
\def\out{{\rm out}}
\def\in{{\rm in}}
\def\mean{{\rm mean}}
\def\zrec{z_{\rm rec}}
\def\zreio{z_{\rm reion}}
\def\wmap{{\it WMAP}} 
\def\planck{{\it Planck}}

\section{Introduction}
The departure of the CMB from statistical isotropy and homogeneity has attracted very serious attention since the first publications of the WMAP data \cite{WMAP1, WMAP3, WMAP5}. Investigations of the low multipole anomalies of the WMAP co-added map and the ILC whole sky map can clarify possible sources of ``contamination" \cite{Jkim1, Chiang03, Coles, Vielva, Eriksen2, Eriksen1, Larson, Oliveira-Costa1, Park1, Copi1, Land1, Coldspot1, Copi2, Schwarz1, Hoftuft1}. The low multipole non-Gaussian features and departure from statistical anisotropy of the CMB can be related to the foreground residuals \cite{Burigana, Naselsky2, Naselsky1, Naselsky3}, systematic effects \cite{Hansen1}, or it could even have a primordial origin, for instance by a violation of the Copernican principle \cite{Odd_C}, a primordial magnetic field \cite{PMF_anomaly, alfven}, or by a non-trivial topology of the Universe \cite{low_quadrupole, spherical_tessellation, Land_Bianchi}. Usually, discussing uncounted residuals of the foregrounds as a source of low multipole anomalies, one can assume that most of those residuals are associated with the Galactic plane, while for instance, the observed quadrupole -octupole alignment \cite{Copi1} is connected with the ecliptic plane. Spergel et al. \cite{spergel} have pointed out, that low multipole anomalies, like a power asymmetry and statistical anisotropy of the CMB signal, correlates with the ecliptic North and South poles, and more generally, reflect the morphology of the map of number of observations (MNO). Indeed, \cite{groeneboom1, groeneboom2} building on the work of \cite{hanlew} and \cite{ack} have identified a quadrupolar power asymmetry, confirmed by the WMAP team \cite{WMAP7:anomaly}. Most likely, this anomaly indicates an influence of the beam asymmetry on the CMB signal, discussed in \cite{hlc}. However, as it was pointed out by \cite{wehus}, in spite of significant distortions of the phases of the CMB signal, an anisotropic beam could only change the beam window function with less then $0.6\%$ at $l\gg 400$, and it is insignificant for estimation of the CMB power spectrum at the low multipole range $l\ll 200$, with changes at the level of $<0.1\%$.\\
Most of these tests are based on the CMB maps, meaning that a high quality of data reduction  is essential. In contrast, 
as recently proposed in \cite{Jkim1, Jkim2}, the parity test requires only the CMB power spectrum, which is usually estimated
with significantly better accuracy than the CMB map \cite{Odd_bolpol}.\\
The idea of the parity test is based on the analysis of the ratio $g(l)$ of the powers, stored in even and odd multipoles. 
For a statistically homogeneous and isotropic random CMB signal, $g(l)$ should have no preferences in respect to $l=even$ or $l=odd$ (see \cite{Kim2011, Parity_review} for discussions).\\
In this paper we would like to focus on an extension of the parity test, quadrupole-octupole alignment and lack of angular correlations at $\theta\ge 60^o$ \footnote{As it was shown in \cite{Kim2011} the anomaly of the correlation function can be explained as manifestation of the parity asymmetry of the CMB power} for the whole sky CMB map, taking under 
consideration the possible contributions from ``new foregrounds" like the Kuiper Belt objects KBO, (see for review \cite{Dikarev}), counting them as a source of contamination of the CMB \cite{Burigana}. Under the standard assumption that the KBOs is localized mainly in the Ecliptic plane, this foreground is potentially "dangerous" for the quadrupole-octupole alignment. More importantly, due to the high degree of symmetry, the morphology of the KBO signal could be remarkably similar to the morphology of the MNO, for instance at the V-band, and at least at the low multipole range  $l\le 20-30$, where the power is comparable to the CMB power spectrum.\\
In \cite{Lew}, the concept of dipole modulation of the CMB is presented. The dipole modulation is effectively an addition of a term in harmonic space, with an amplitude $A$. This extra dipole term multiplies the primordial (true) cmb temperature value, at a given location on the sphere, with a number between $-A$ and $A$. If the location coincides with the direction of the dipole modulation, the number is $A$, and if opposite the dipole modulation direction, the number is $-A$. Effectively this introduces a dipole component in the measured map. See \cite{Lew} for further details. The critical point of our analysis is the assumption, that the dipole modulation of the CMB, is associated with an unknown systematic effect. Thus it affects not only the measured primordial CMB signal, but all the measurements of the foregrounds, including the KBO, as well.\\
We estimate the temperature change which the model-KBOs contribute to the CMB. We show that unlike the model of \cite{Burigana}, de-correlation of the ILC 7 and KBO leads to a decrease of the power of even and odd harmonics, improving the shape of the parity parameter $g(l)$.\\
Moreover, we will show that a KBO-ILC 7 de-correlation could change the significance of the quadrupole-octupole alignment, down to the level of spontaneous (chance) correlations. Thus, if the KBOs are responsible for the parity asymmetry of the CMB, one should be cautious of the alignment of quadrupole and octupole components of the CMB.\\
The outline of the paper is the following. In section 2 we introduce the basic characteristics of the parity asymmetry. In section 3, we discuss the model of the Kuiper Belt. Then, in section 4, we present a method of cross-correlations of the KBO foreground and the WMAP ILC 7 map at low multipole range. Section 5 is devoted to the analysis of general properties of the cross-correlations between ILC 7 and KBO signal. In section 6 we investigates the distortion of even and odd component of the ILC 7 map. Section 7 is devoted to the cleaning of the ILC 7 map by de-correlation with KBO emission. Additionally we look at the implications for the parity asymmetry and the quadrupole-octupole alignment, and test various properties of the KBO filtered signal in section 8. Finally, in section 9, we summarize our results and draw our conclusions.

\section{Odd multipole preference of the power spectrum}
The temperature fluctuations of the CMB can be decomposed into spherical harmonics in the standard way:
\begin{eqnarray}\label{a_lm}
\Delta T (\theta,\phi) = \sum ^{\infty}_{l=1} \sum ^{l}_{m=-l} a_{l,m} Y_{l,m}(\theta,\phi),
\end{eqnarray}
where $a_{l,m}$ is the coefficient of decomposition: $a_{l,m} = |a_{l,m}| \exp(i \phi_{l,m})$, with $\phi_{l,m}$ as the phase.\\
Under the assumption of total Gaussian randomness, as predicted by the simplest inflationary model, the amplitudes $|a_{l,m}|$ are distributed according to Rayleigh's Probability Density Function (PDF) and the phases of $a_{l,m}$ are supposed to be evenly distributed in the range $[0,2 \pi]$ \cite{Bardeen1986}.\\
For any signal $T(\hat{\mathbf n})$ defined on the sphere, one can extract a symmetric $ T^+(\hat{\mathbf n})=T^+(-\hat{\mathbf n})$ and an anti-symmetric $ T^-(\hat{\mathbf n}) =- T^-(-\hat{\mathbf n})$ component, where
\begin{eqnarray} 
T^{\pm}(\hat{\mathbf n})&=&\sum ^{\infty}_{l=1} \sum_{m=-l}^{l}a_{l,m}\Gamma^{\pm}(l)Y_{l,m}(\hat{\mathbf n}),
\label{gam}
\end{eqnarray}
and $\Gamma^{+}(l)=\cos^2(\frac{\pi l}{2})$, $\Gamma^{-}(l)=\sin^2(\frac{\pi l}{2})$, $Y_{l,m}(\hat{\mathbf n})=(-1)^l\,Y_{l,m}(-\hat{\mathbf n})$.\\
For the concordance $\Lambda$CDM cosmological model with initial Gaussian adiabatic perturbations, we do not expect any features distinct between even and odd multipoles. However, there have been reported power contrast between even and odd multipoles of WMAP $TT$ power spectrum \cite{Jkim1, Jkim2, WMAP7:anomaly, Odd_bolpol, Universe_odd}. The corresponding estimator for "even and odd" asymmetry of the CMB power spectrum is given by \cite{Jkim1}:
\begin{eqnarray}
g(l)=\frac{\sum_{n_{min}}^l n(n+1)C(n)\Gamma^+(n)}{\sum_{n_{min}}^l n(n+1)C(n)\Gamma^-(n)}
\label{par_est}
\end{eqnarray}
where $C(n)$ is the usual power spectrum.
At lowest multipoles ($2\le l\le 23$), there is odd multipole preference (i.e. a power excess in odd and a deficit in even multipoles) as discussed in \cite{Jkim1, Jkim2, Odd_bolpol, Universe_odd}.

\FIGURE{
 \centerline{\includegraphics[scale=0.15]{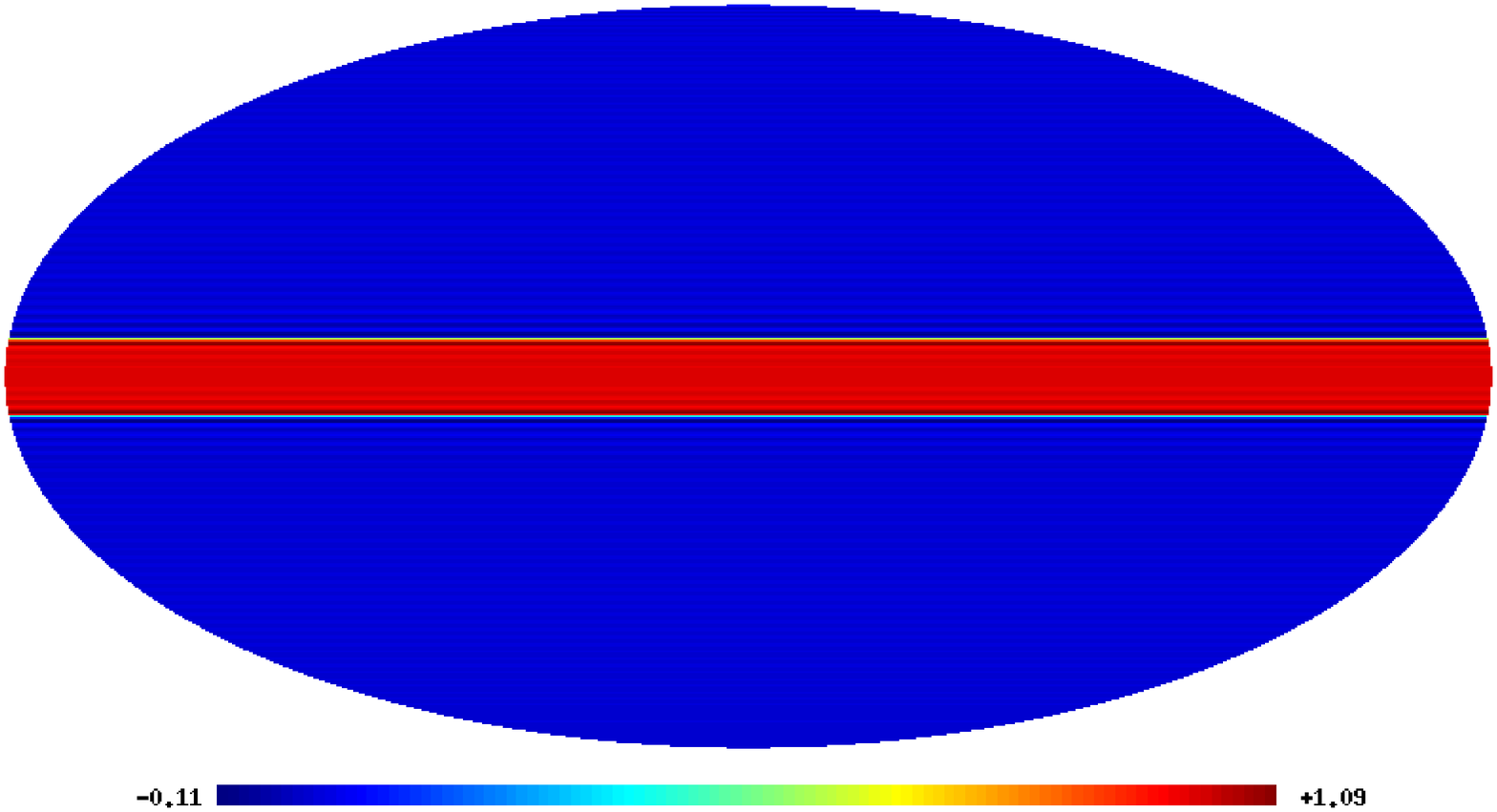}
\includegraphics[scale=0.15]{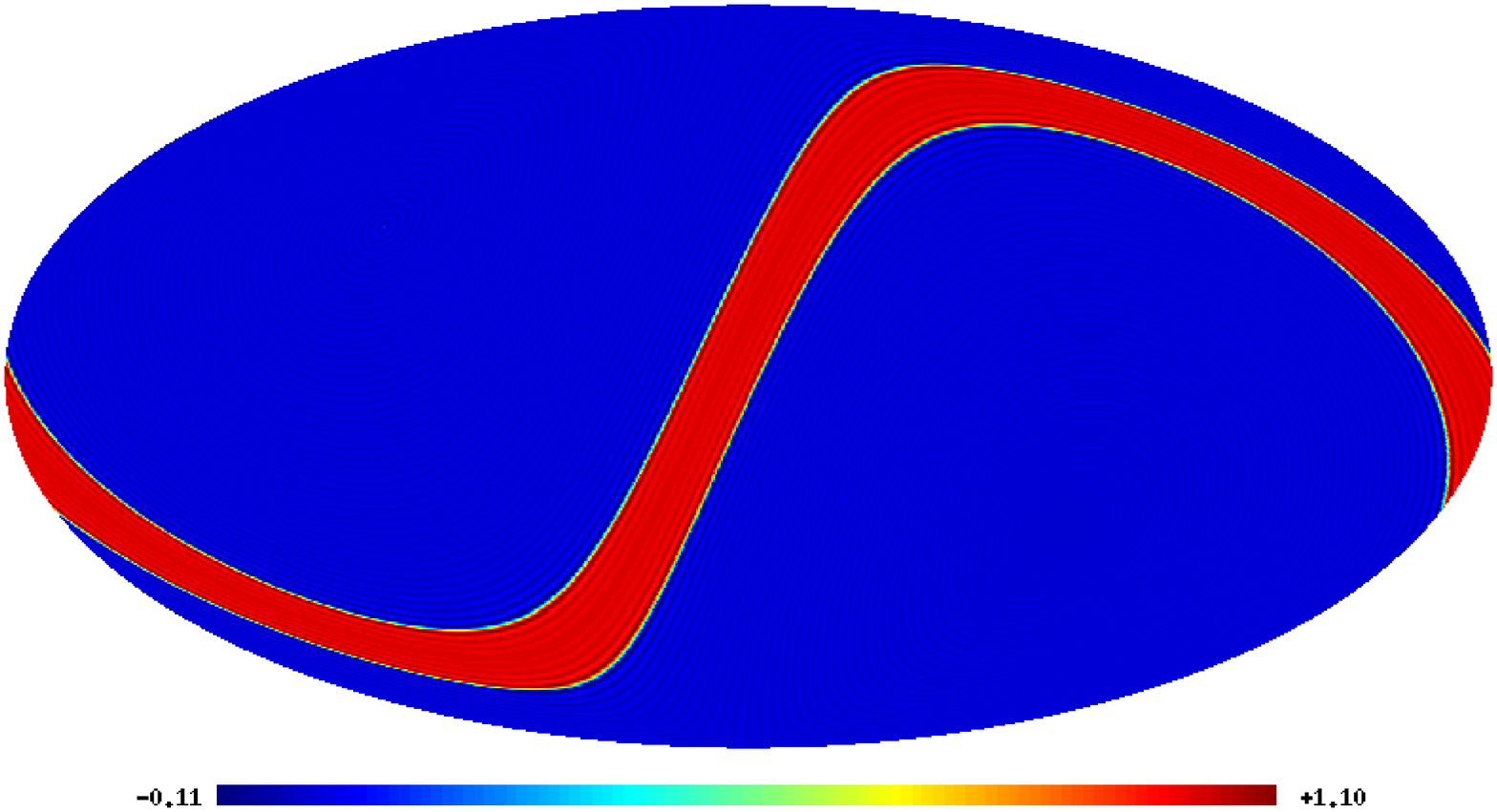}}
 \centerline{\includegraphics[scale=0.15]{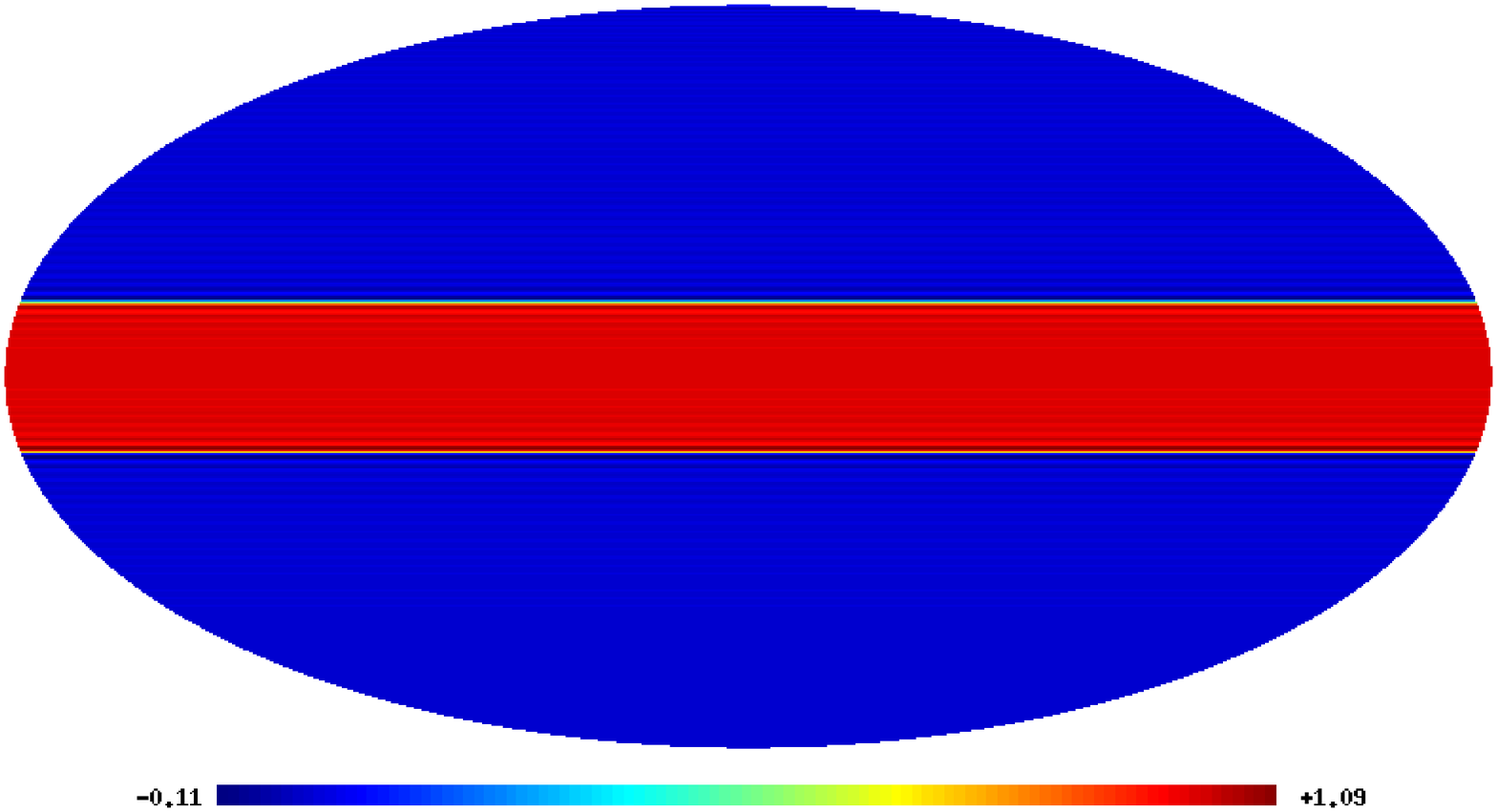}
\includegraphics[scale=0.15]{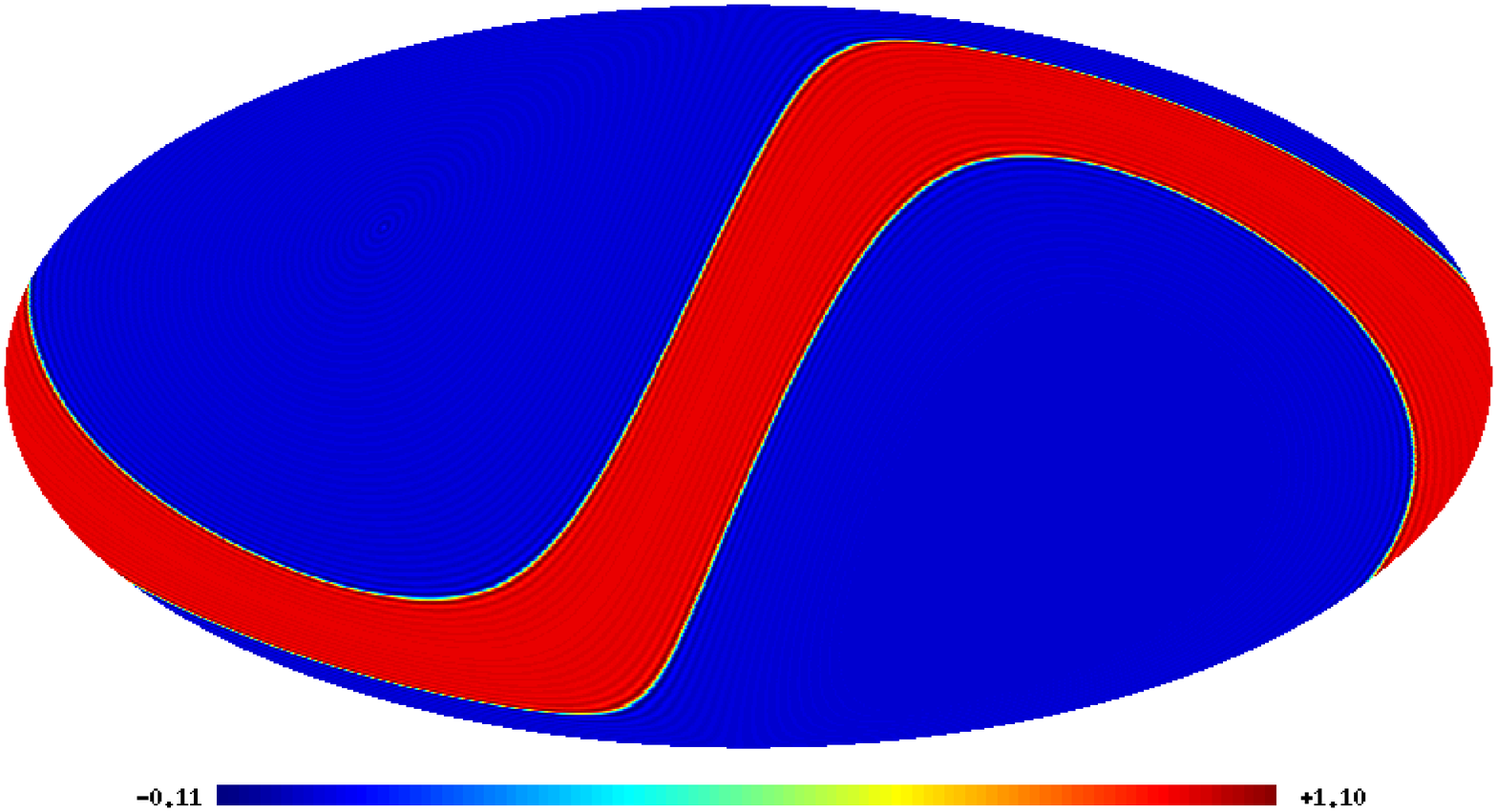}}
 \centerline{\includegraphics[scale=0.15]{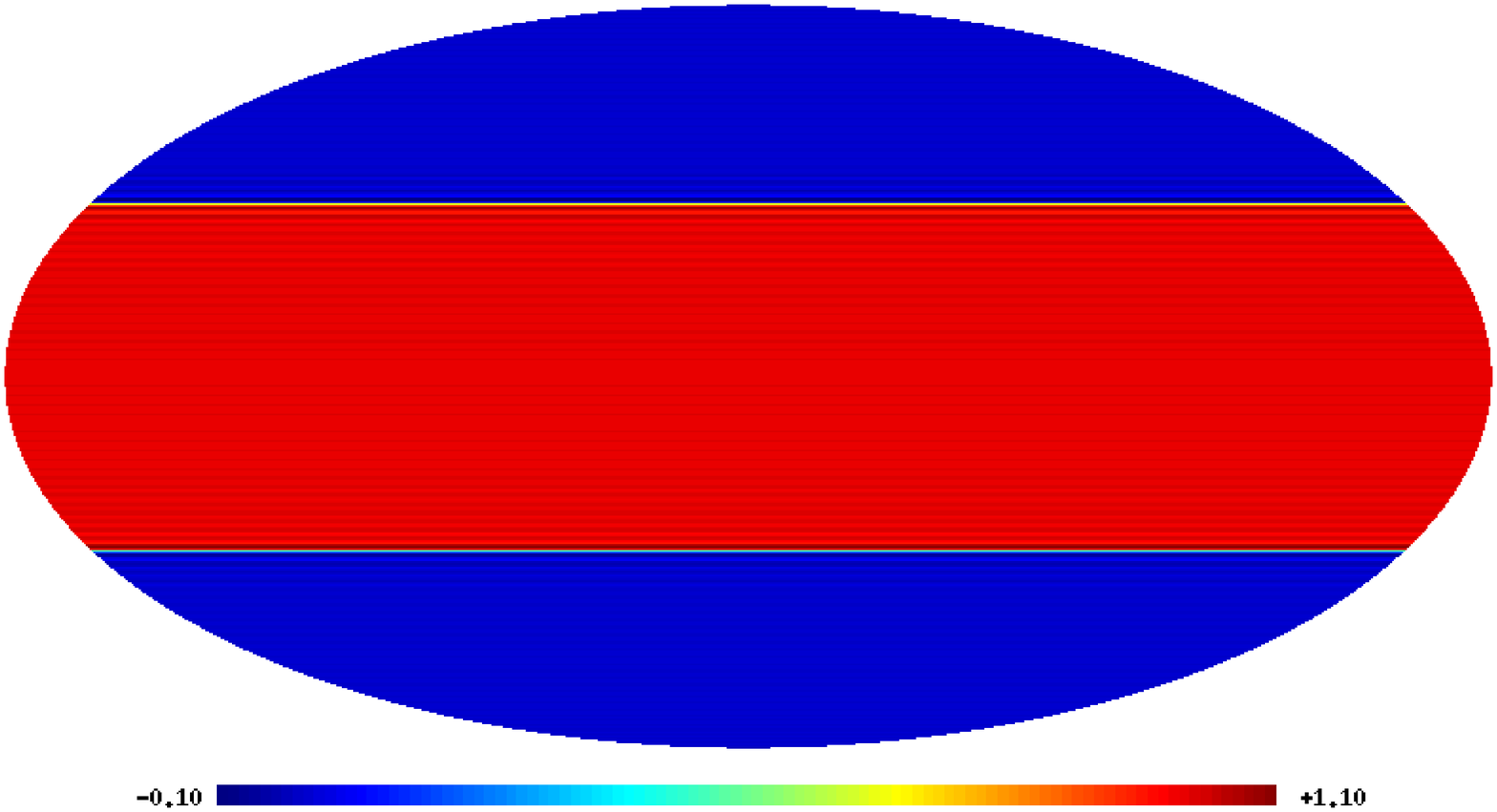}
\includegraphics[scale=0.15]{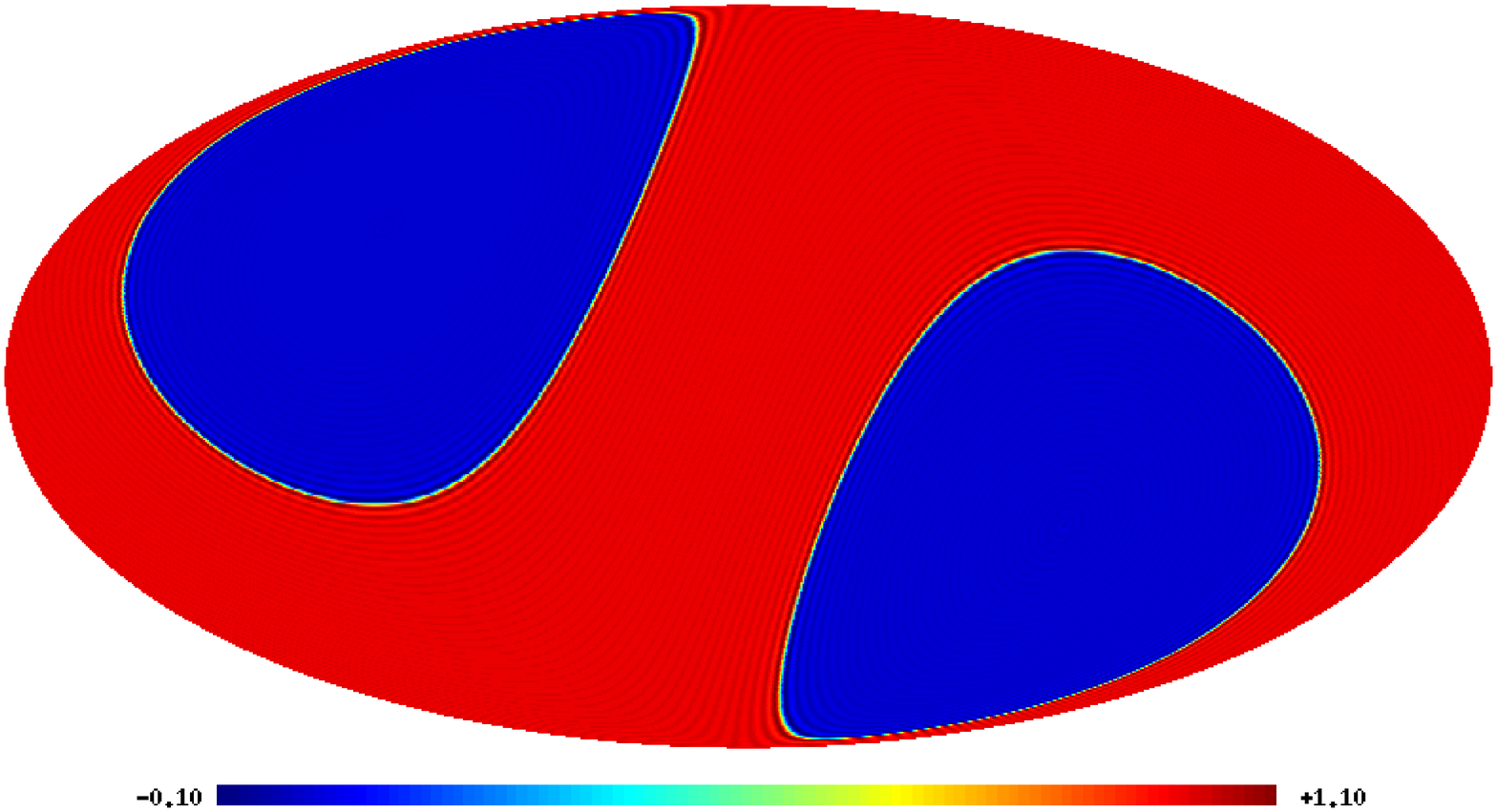}
}
\caption{Left column. Model for the KBO emissivity distribution (arbitrary units) in Ecliptic coordinates for  $H=15^o$ (top), for $H=30^o$ (the middle panel) and $H=70^o$(the  bottom panel). Right column show the KBO emissivity in the Galactic coordinates. All maps carry equal power.
}
\label{kbo} }

\section{KBO's as a 'new foreground'}
The contribution from the residuals of the foregrounds into parity asymmetry of the CMB was widely discussed in \cite{Burigana, Jkim1, Universe_odd}. Unlike the first two papers, the last one claims that uncounted foregrounds, like emissivity of dust from the Kuiper Belt, could explain the detected parity asymmetry of the CMB. The main idea of the method proposed in \cite{Burigana} is that a very symmetric foreground, when taken into account and subtracted from the WMAP whole-sky ILC map, could amplify the CMB power stored in even multipoles, increasing the parity parameter $g(l)$ (i.e. mitigating the observed parity asymmetry) for the multipole range $2\le l\le 23$. An illustration of the idea is shown in Fig. \ref{kbo}, where the models of the KBO emissivity are shown in Galactic and Ecliptic coordinates for different angular heights of the KBO $H=15^o,30^o$ and $70^o$.\\
The blackbody-like radiation from the KBO affects the intensity of the microwave sky in the optically thin limit as follows:
\begin{eqnarray}
I(\nu,\hat{\mathbf r})=(B(\nu,T_{\mathrm{KBO}})-B(\nu,T_{\mathrm{CMB}}))\,\tau(\hat{\mathbf r}), \label{I_model}
\end{eqnarray}
where $\nu$ is the observation frequency and $B(\nu,T)$ is the blackbody radiation spectrum at temperature $T$, and $\tau(\hat{\mathbf r})$ is the optical depth of KBOs at the sky direction $\mathbf r$. The second term on the right hand side of Eq.(\ref{I_model}) arise from the occultation of CMB photons by the KBO. The FIRAS data imposed the most stringent constraints on the sky-averaged optical depth of KBOs, which is $\tau \lsim 3\times 10^{-7}$ \cite{Burigana, Fixen:dipole, Bennett:dipole, Mather:CMB_T_1999, TNO_Babich}.\\
The heating is mainly due to the radiation from the Sun. Therefore, it is possible to calculate the temperature of these objects at the equilibrium, assuming that it arises from the conversion of the solar radiation absorbed by the object into microwave emission. At a distance of 40AU, where KBOs are most densely populated, we find an equilibrium temperatures of $\sim 43.7$K \cite{TNO_Babich, KBO_l50}. In Table \ref{f}, we show the frequency spectrum of KBO divided by the CMB anisotropy spectrum $f(\nu)$.
\TABLE{
\caption{Frequency spectrum of KBOs at WMAP frequencies (in GHz) normalized to CMB anisotropy spectrum.}
\begin{tabular}{cccccc}
\hline
band &  K& Ka& Q& V& W\\
\hline
$f(\nu)$& 41.48 & 42.07 &42.67 &44.8 &50.42\\
\hline
\end{tabular}
\label{f} }
As evident, the frequency spectrum of KBOs does not vary significantly even between the both ends of the WMAP observation frequency:
\begin{eqnarray}
(f(\nu_K)-f(\nu_W))/f(\nu_K)\lsim 0.22,
\end{eqnarray}
Therefore, the KBO emission may be confused with intrinsic CMB anisotropy, even when the contribution to the microwave sky emission is sizable. For the allowed values of the optical depth, $\tau\lsim 3\times 10^{-7}$, the KBOs may have an averaged effect on the CMB data as big as $\sim 15\,\mu$K. Provided the KBOs have certain large-scale patterns, KBOs may have an effect large enough to be the culprits of reported large-scale CMB anomalies \cite{Jkim1, Jkim2, Oliveira-Costa1,  Copi1, Land1, Copi2, Schwarz1, Odd_C, Odd_bolpol, Universe_odd, multipolevectors, Copi3, Land2, Lowl_anomalies, Odd_tension}. 

\subsection{Angular distribution of the KBO emissivity}
To investigate  the emissivity of  Kuiper Belt, we adopt the models by \cite{Burigana}, where the Kuiper Belt covers a symmetrical band in the Ecliptic plane of constant disk height $H$, with uniform temperature distribution. Following \cite{Burigana}, we pick three models, defined by the angle $\pm H_{KBO}/2$ from the Ecliptic plane, where we choose $ H_{KBO} = 15^{\circ}, 30^{\circ}$ and $70^{\circ}$ respectively.\\
For the three values of $H_{KBO}$, we set the temperature inside the Kuiper Belt, so that it agrees with the upper bound of $15 \mu$K for the entire sky, as calculated previously in the article. In practice, we do this by calculating the respective fractions of the total sky area, for each value of $H_{KBO}$, and finding the required temperature inside the band from this value:
\begin{eqnarray}
T_{kbo}(\theta,\phi)=B\Theta\left[\theta-\frac{1}{2}(\pi-H)\right]\Theta\left[\frac{1}{2}(\pi+H)-\theta\right]
\label{tt}
\end{eqnarray}
where we work in spherical coordinates, with H in radians and where $B$ is a normalization constant and $\Theta(x)$ is the Heaviside-function. In the discussion which follows, we will use the average temperature of the KBO emission as normalization of $B$:
\begin{eqnarray}
T_{KBO}=\frac{1}{4\pi}\int_0^{2\pi}d\phi\int_0^{\pi}d\theta\sin\theta T_{kbo}(\theta,\phi)=B\sin\left(\frac{H}{2}\right).
\label{norm}
\end{eqnarray}
Now, we decompose the KBO signal into spherical harmonics and get the following coefficients of decomposition:
\begin{eqnarray}
f_{l,m}=\sqrt{\frac{2l+1}{4\pi}}B\Gamma^+(l)\delta_{m,0}\int^{(\pi+H)/2}_{(\pi-H)/2}d\theta\sin\theta P_l(\cos\theta)
\label{kbot}
\end{eqnarray}
For $H\ll\pi/2$ in Eq.(\ref{kbot}) we can use the following Taylor series representation: $P_l(x)\simeq P_l(0)+\frac{1}{2}P''_l(0)x^2$, where the two primes denotes the second derivative. The first derivative vanishes for even $l=2n$, due to the symmetry of the model ($\frac{dP_l(x)}{dx}|_{x=0}=0$). This gives 
\begin{eqnarray}
P_l(0)=\sqrt{\pi}\!\left[\Gamma(\frac{l}{2}\!+\!1)\Gamma(\frac{1\!-\!l}{2})\right]^{-1}\!\!,
P''_l(0)=-l(l\!+\!1)P_l(0)
\label{kbot1}
\end{eqnarray}
Here $\Gamma(x)$ is the Gamma-function. For $l(l+1)H^2/24\ll 1$, from Eq.(\ref{kbot}-\ref{kbot1}) one can get:
\begin{eqnarray}
f_{l,0}\simeq\sqrt{\frac{2l+1}{\pi}}B\sin(\frac{H}{2})P_l(0)\Gamma^+(l)\left[1-\frac{l(l+1)}{24}H^2\right]
\label{kbot2}
\end{eqnarray}
Thus, for $l\ll 35 \left(\frac{15^o}{H(deg)}\right)$ the coefficients of the expansion $f_{l,0}$ have the following analytical representation:
\begin{eqnarray}
f_{l,0}\simeq\sqrt{\frac{2l+1}{\pi}}T_{KBO}(-1)^n\frac{(2n-1)!!}{2^nn!},\hspace{0.5cm}l=2n
\label{kbot3}
\end{eqnarray}
In order to test the influence of our models on the lowest multipoles, we now find the power spectrum for each value of $H_{KBO}$ and compare them with the ILC power spectrum. The results are presented in Fig. \ref{alm_pat}.
\FIGURE{
 \centerline{\includegraphics[scale=0.8]{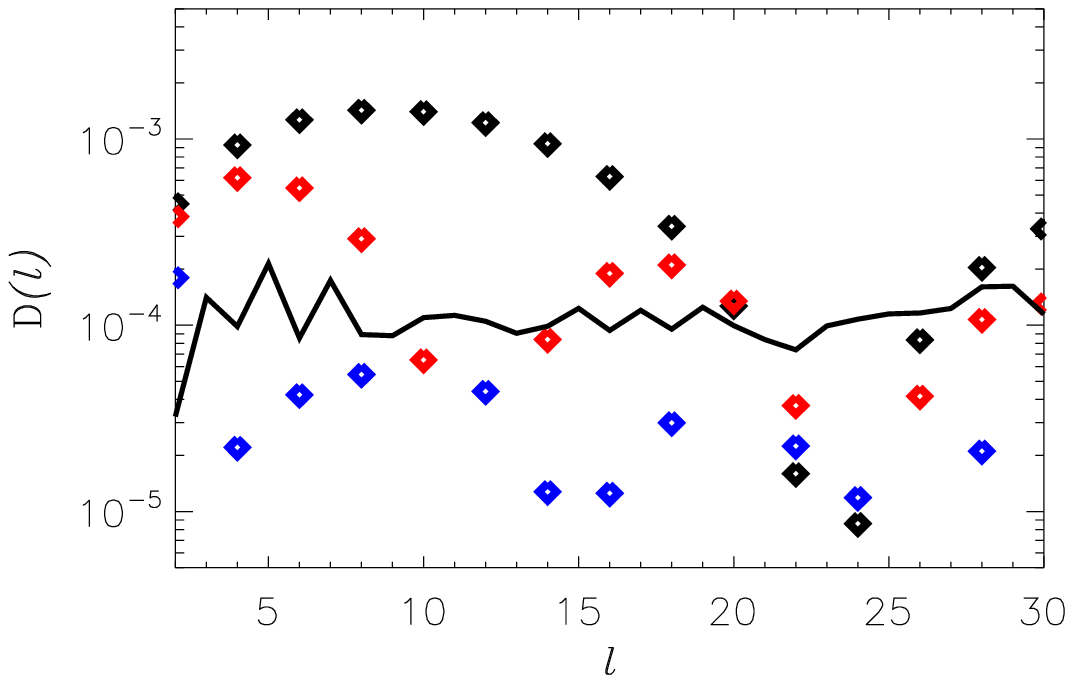}}
\caption{Power spectrum $D(l)=\frac{l(l+1)}{2\pi}C(l)$ (in units of $mK$) for the three KBO models, with $H_{KBO} = 15^{\circ}$ (black) $30^{\circ}$ (red) and $70^{\circ}$ (blue), compared to the ILC power spectrum (black line).
}
\label{alm_pat} }
It is clear, that for all 3 values of $H_{KBO}$, the power of the KBO quadrupole is strong enough to affect the ILC quadrupole. For higher multipoles, $H_{KBO} = 15^{\circ}$ can affect the even multipoles up to $l=19$.

\section{CMB-KBO cross-correlation}
In this section we will discuss general properties of the KBO foreground, basing our analysis on the symmetry in the Ecliptic system of coordinates. Firstly, we will assume that the emissivity of the foreground $S(\theta_e,\phi_e)$ 
is very symmetric with respect to the plane $\theta_e=\pi/2$ in Ecliptic coordinates, and do not depend on the azimuthal coordinate $\phi_e$: $S(\theta_e,\phi_e)=S(\theta_e)$. This assumption, leads to the following properties of the foreground: $f_{l,m}=f(l)\delta_{m,0}\delta_{l,2n}$, $n=1,2..$., clearly demonstrating that only even multipoles can be affected by the KBO-foreground.\\
At the same time, at the level of $3\sigma$ \cite{Lew}, the ILC 7 map has dipole modulation, which will provide a coupling between the even and odd components of the KBO emission, depending on the orientation of the dipole term and the corresponding amplitude of modulation. Here we assume that this kind of modulation has a non-cosmological (systematic) origin, leading to a distortion of the CMB and the foregrounds as well.\\
To assess the problem of a possible contamination of the ILC by the KBO-foreground, we will use the model of the CMB signal for a given direction on the sky $\textbf{n}$:
\begin{eqnarray}
T_c({\textbf{n}})
&=&T_{ilc}({\textbf{n}})- \left[1+\Upsilon({\textbf{q}},{\textbf{n}})\right]\chi_{kbo}({\textbf{n}}) \nonumber \\
&=&T_{ilc}({\textbf{n}})- \chi_{kbo}({\textbf{n}}) - \varepsilon({\textbf{n}})
\label{res}
\end{eqnarray}
where $T_c({\textbf{n}})$ and $T_{ilc}({\textbf{n}})$ are the temperature of the intrinsic CMB signal and the ILC signal respectively and $\chi_{kbo}({\textbf{n}})\propto T_{kbo}({\textbf{n}})$ corresponds to the residuals of the KBO emissivity. $\Upsilon({\textbf{q}},{\textbf{n}})$ is the function of dipole modulation, dependent on the given sky direction $\textbf{n}$, and the direction of dipole modulation $\textbf{q}$. Finally $\varepsilon({\textbf{n}})=\Upsilon({\textbf{q}},{\textbf{n}})\chi_{kbo}({\textbf{n}})$ is the dipole-modulated component of the KBO signal.\\
Let us briefly mention some possible causes for this dipole modulation. Bear in mind, that the cause must be systematic in nature, since it must affect both the primordial CMB signal and the foregrounds. One possibility to consider is the effects of the calibration, including striping, during creation of the map \cite{WMAP1_errors}. An incomplete compensation for the effects of striping could give rise to an artificial dipole modulation in the map. A similar possibility arises from the residuals of the model for the gain factor, where an inconsistency between model and actual gain, could result in a dipole modulation \cite{WMAP3}.\\
In the forthcoming discussion we will assume that the function of dipole modulation takes the simple form $\Upsilon({\textbf{q}},{\textbf{n}})=A({\textbf{q}}{\textbf{n}})$, where $A=const$ is the dipole modulation amplitude, and ${\textbf{q}}{\textbf{n}}$ is a dot product of the two vectors $\textbf{q}$ and $\textbf{n}$. We would like to point out that the amplitude of the dipole modulation can be estimated from Fig. \ref{alm_pat}, with $A\simeq 0.3-0.4$, which is comparable with the results by \cite{Lew} at the $\le2\sigma$-level.\\
As a result of the model above, the residuals of the KBO emissivity (the second term in Eq.(\ref{res})), are symmetric with respect to inversion: $ \chi_{kbo}({\textbf{n}})=\chi_{kbo}(-\textbf{n})$, while the last term, dependent on the dipole, is anti-symmetric: $\varepsilon({\textbf{n}})=-\varepsilon(-{\textbf{n}})$. In the multipole domain these two component contribute to the coefficients of the spherical harmonic decomposition additively:
\begin{eqnarray}
c_{l,m}=a_{l,m}-\varsigma(l)\chi_{l,m}\Gamma^+(l)-\rho(l)\varepsilon_{l,m}\Gamma^-(l)=a_{l,m}-\Pi_{l,m}
\label{res1}
\end{eqnarray}
where $c_{l,m}$ is the primordial coefficient of decomposition, $\Gamma^{\pm}$ are given by Eq.(\ref{gam}), $\varsigma(l)$ and $\rho(l)$ are weighting coefficients, and $\varepsilon_{lm}$ can be expressed as
\begin{eqnarray}
\varepsilon_{lm}=\sum_{m'=-1}^{1}\sum_{l''=0}^{\infty}\sum_{m''=-l''}^{l''}  (-1)^{m} \left(b_{1m'}f_{l''m''}\right)\nonumber\\
\sqrt{\frac{3(2l''+1)(2l+1)}{4\pi}}
\left(\begin{array}{ccc}l'' & 1 & l \\ 0 & 0 &0 \end{array}\right)
\left(\begin{array}{ccc}l'' & 1 & l \\ m'' & m' & -m\end{array}\right).
\label{varep}
\end{eqnarray}
Here $b_{lm}$ relates to the parameters $A$ and $\mathbf n$ by the relation $b_{lm}=\frac{4\pi A}{3} Y_{lm}^*(\mathbf n)$.\\
As stated earlier, for the given model of the KBO we have $f_{l,m}=f(l)\Gamma^+(l)\delta_{m,0}$, and thus, in Ecliptic coordinates, Eq.(\ref{varep}) can be simplified as follows:
\begin{eqnarray}
\varepsilon_{lm}&=&\sum_{m'=-1}^{1}\sum_{l''=2}^{\infty}(-1)^{m}\Gamma^+(l'')b_{1m'}f_{l''}\sqrt{\frac{3(2l''+1)(2l+1)}{4\pi}}\nonumber\\
&\times&\left(\begin{array}{ccc}l'' & 1 & l \\ 0 & 0 &0 \end{array}\right)
        \left(\begin{array}{ccc}l'' & 1 & l \\ 0 & m' & -m\end{array}\right).
\label{varep1}
\end{eqnarray}

\section{General remarks about KBO -intrinsic CMB cross-correlation} \label{section:Cross_C}
In this section we would like to show, that excluding dipole modulation, the primordial CMB signal need a very strong cross-correlations with the uncounted foregrounds. The case is entirely different with the inclusion of dipole modulation however.
Let us discuss this issue in more detail. Following Eq.(\ref{res1}), we assume that the ILC 7 signal can be represented in the following way
\begin{eqnarray}
a_{l,m}=c_{l,m}+{\Pi}_{l,m}
\label{for}
\end{eqnarray}
Let us define the coefficients of cross-correlations between the ILC map and the foregrounds, and the primordial CMB and the foregrounds as follows:
\begin{eqnarray}
K(l)=\frac{\sum_m\left[a_{l,m}\Pi^*_{l,m}+a^*_{l,m}\Pi_{l,m}\right]}{2\left(\sum_m|a_{l,m}|^2\sum_{m'}|\Pi_{l,m'}|^2\right)^{\frac{1}{2}}},\nonumber\\
\kappa(l)=\frac{\sum_m\left[c_{l,m}\Pi^*_{l,m}+c^*_{l,m}\Pi_{l,m}\right]}{2\left(\sum_m|c_{l,m}|^2\sum_{m'}|\Pi_{l,m'}|^2\right)^{\frac{1}{2}}}
\label{cros}
\end{eqnarray}
Combining Eq.(\ref{for}) and Eq.(\ref{cros}) we can get the following equation for the $\varsigma(l)$ -parameter:
\begin{eqnarray}
\varsigma(l)=K(l)\left(\frac{\sum_m|a_{l,m}|^2}{\sum_m|\Pi_{l,m}|^2}\right)^{\frac{1}{2}}-\kappa(l)\left(\frac{\sum_m|c_{l,m}|^2}{\sum_m|\Pi_{l,m}|^2}\right)^{\frac{1}{2}}
\label{alpha}
\end{eqnarray}
So, as one can see from Eq.(\ref{alpha}), the coefficient $\varsigma(l)$ play a role of effective coefficient of
cross-correlations between ILC, intrinsic CMB and the foreground residuals. From Eq.(\ref{for}) and Eq.(\ref{alpha})
one can define the power spectrum of the primordial CMB as:
\begin{eqnarray}
C_p(l)=\frac{1}{2l+1}\sum_m|c_{l,m}|^2=V(l)C_{ilc}(l)
\label{pow}
\end{eqnarray}
where $C_{ilc}(l)$ is the power of the ILC 7 map, and
\begin{eqnarray}
V(l)=\frac{1-K^2(l)}{1-\kappa^2(l)}
\label{v}
\end{eqnarray}
is the factor of modulation of the ILC 7 power. Now, by using Eq.(\ref{pow}) we can estimate the parity parameter for the primordial CMB as 
\begin{eqnarray}
g_p(l)=\frac{\sum_{n=2}^l n(n+1)V(n)\Gamma^+(n)C_{ilc}(n)}{\sum_{n=2}^l n(n+1)V(n)\Gamma^-(n)C_{ilc}(n)}
\label{par_new}
\end{eqnarray}
In order to increase the contribution of even multipoles in parity parameter $g_p(l)$, the functions $V^+(l)=V(l)\Gamma^+(l)$ and $V^-(l)=V(l)\Gamma^-(l)$ should satisfy the following conditions: $V^+(l)\gg V^-(l)$. That means that there are only two variants: $|\kappa^+|\rightarrow 1$ or $ |K^-|\rightarrow 1$, where the $+$ and $-$ in the superscript symbolizes that the $\kappa$ or the $K$ is drawn from either $V^+$ or $V^-$ respectively. The first case (with $|\kappa^+|\rightarrow 1$ and symmetric foreground) was discussed in \cite{Burigana}, and it requires a significant coupling between the primordial CMB and KBO-foreground at the level $|\kappa^+|\sim 0.7-1$. \\
Due to the very clear established non-Gaussian properties of the KBO-foreground, a high correlation between it and the primordial CMB temperature anisotropy, is in contradiction with the assumption about a statistical isotropic and Gaussian intrinsic CMB signal. Thus, the improvement of the parity parameter by an increase of $|\kappa^+|$ (hereafter \textbf{\textit{Model 1}}) will have lost its theoretical basis.\\
In contrast, for $|\kappa^{\pm}|\ll 1$, the only way to increase $g_p(l)$ from Eq.(\ref{par_new}), keeping the assumption about Gaussianity and statistical isotropy of the intrinsic CMB signal, is to get $ |K^-|\rightarrow 1$ and $|K^+|\le 1$ (\textbf{\textit{Model 2}}). Thus in the optimal case, we have a high correlation between the ILC signal and the KBO-foreground for odd multipoles, and at the same time, a relatively low correlation between the ILC signal and the KBO-foreground for even multipoles. At the end of section \ref{section:l=5}, we will return to these conditions.

\section{Distortion of odd multipoles. Estimation of $K^-$ in Model 2}
\TABLE{
\caption{$a_{3,m}$ coefficients of the ILC 7 octupole in the Ecliptic coordinates.}
\begin{tabular}{cccc}
\hline
m & $\Re e(a_{3,m})$ & $\Im m (a_{3,m})$ & $|a_{3,m}|$ \\
\hline
0 & $-3.9242 \cdot 10^{-2}$ & 0 & $3.9242 \cdot 10^{-2}$ \\
\hline
1 & $6.3435 \cdot 10^{-3}$ & $1.9066 \cdot 10^{-2}$ & $2.0093 \cdot 10^{-2}$ \\
\hline
2 & $-2.0306 \cdot 10^{-2}$ & $-1.7310 \cdot 10^{-2}$ & $2.6681 \cdot 10^{-2}$ \\
\hline
3 & $-5.4958 \cdot 10^{-3}$ & $-2.2534 \cdot 10^{-3}$ & $5.9398 \cdot 10^{-3}$ \\
\hline
\end{tabular}
\label{alm} }
The contribution from the KBO emissivity into the coefficient of cross-correlations $K^-=K(l)\Gamma^-(l)$ is given by Eq.(\ref{for}) and Eq.(\ref{cros}):
\begin{eqnarray}
K^-(l)=\frac{\sum_m\left[a_{l,m}\varepsilon^*_{l,m}+a^*_{l,m}\varepsilon_{l,m}\right]}{2\left(\sum_m|a_{l,m}|^2\sum_{m'}|\varepsilon_{l,m'}|^2\right)^{\frac{1}{2}}},
\label{kmin}
\end{eqnarray}
where
\begin{eqnarray}
\varepsilon_{l,m}&=& A\cos\Theta[\alpha_1(l,m)f_{l+1,m}+\beta_1(l,m)f_{l-1,m}] \nonumber\\
&+& (A/2)\sin\Theta ~e^{i\Phi}[\alpha_2(l,m)f_{l-1,m+1}-\beta_2(l,m)f_{l+1,m+1}] \nonumber\\
&+& (A/2)\sin\Theta ~e^{-i\Phi}[\alpha_3(l,m)f_{l+1,m-1} - \beta_3(l,m)f_{l-1,m-1}],
\label{vareps}
\end{eqnarray}
and
\begin{eqnarray}
\alpha_1(l,m)&=&\left(\frac{(l+m+1)(l-m+1)}{(2l+1)(2l+3)}\right)^{1/2},\nonumber\\
\beta_1(l,m)&=&\left(\frac{(l+m)(l-m)}{(2l+1)(2l-1)}\right)^{1/2},\nonumber \\
\alpha_2(l,m)&=&\left(\frac{(l-m)(l-m-1)}{(2l+1)(2l-1)}\right)^{1/2},\nonumber\\
\beta_2(l,m)&=&\left(\frac{(l+m+1)(l+m+2)}{(2l+1)(2l+3)}\right)^{1/2},\nonumber \\
\alpha_3(l,m)&=&\left(\frac{(l-m+1)(l-m+2)}{(2l+1)(2l+3)}\right)^{1/2},\nonumber\\
\beta_3(l,m)&=&\left(\frac{(l+m-1)(l+m)}{(2l+1)(2l-1)}\right)^{1/2}.
\label{vareps1}
\end{eqnarray}
Here $\Theta$ and $\Phi$ are the coordinates of the dipole modulation in the Ecliptic system of coordinates, and $f_{l,m}=f_l\Gamma^+(l)\delta_{m,0}$ is the KBO foreground. Due to the high symmetry of the KBO foreground, only the even multipoles $l=2n, ~n=1,2,..$ are presented in Eq.(\ref{vareps}-\ref{vareps1}). That means that $\varepsilon_{l,m}$-term in Eq.(\ref{vareps}) has only three non-vanishing component with odd $l=2n+1$ and $m=0,\pm 1$. Namely, 
\begin{eqnarray}
\varepsilon_{2n+1,0}&=&A\cos\Theta[\alpha_1(2n+1,0)f_{2n+2} + \beta_1(2n+1,0)f_{2n}],\nonumber\\
\varepsilon_{2n+1,1}&=&(A/2)\sin\Theta~e^{-i\Phi}[\alpha_3(2n+1,1)f_{2n+2} - \beta_3(2n+1,1)f_{2n}],\nonumber\\
\varepsilon_{2n+1,-1}&=&(A/2)\sin\Theta ~e^{i\Phi}[\alpha_2(2n+1,-1)f_{2n} - \beta_2(2n+1,-1)f_{2n+2}]
\label{vareps2}
\end{eqnarray}
where: $l=2n+1$, and
\begin{eqnarray}
\alpha^2_3(l,1)=\beta^2_2(l,-1)=\frac{l(l+1)}{(2l+1)(2l+3)},\nonumber\\
\beta^2_3(l,1)=\alpha^2_2(l,-1)=\frac{l(l+1)}{(2l+1)(2l-1)},\nonumber\\
\alpha^2_1(l,0)=\frac{(l+1)^2}{(2l+1)(2l+3)},\nonumber\\
\beta^2_1(l,0)=\frac{l^2}{(2l+1)(2l-1)}.
\label{vareps3}
\end{eqnarray}

\subsection{Particular model of dipole modulation. Maximization of the octupole component of cross-correlation} 
As it is follows from Eq.(\ref{kmin}) and Eq.(\ref{vareps}), the coefficient of cross-correlations depends on orientation of the dipole in the Ecliptic coordinates ($\Theta,\Phi$). Let us discuss one particular choice of $\Theta$ and $\Phi$, which maximize the coefficient of cross-correlations $K^-(l=3)$. As we have pointed out in introduction, our main goal is to show that KBO foreground can resolve the parity problem at the range of multipoles $2\le l\le 23$. However, this model can potentially explain the quadrupole-octupole alignment also, due to coupling between even and odd components of $\varepsilon_{l,m}$, given by Eq.(\ref{vareps2}). In particular, the octupole component of $\varepsilon_{l,m}$, according to Eq.(\ref{vareps2}) depends on the linear combination of the quadrupole and $l=4$-components of $f_{l,m}$. Due to the azimuthal symmetry of the KBO emissivity, these components have opposite phases. Thus, the coefficient of the KBO quadrupole and octupole cross-correlation depends only on $\Theta$ and $\Phi$. Since KBO quadrupole and octupole components could correlate
with the corresponding components of the ILC, maximizing correlations between the ILC octupole and $\varepsilon_{3,m}$, will provide the optimal way for a dis-alignment of the intrinsic quadrupole and octupole.\\
In Table \ref{alm} we show all the component of the ILC octupole in ecliptic coordinates. As one can see from this Table, the phase of $(3,1)$ component of the octupole is $\varphi_{3,1}=1.227$, which means that in order to maximize the $K^-(l=3)$-coefficient, the azimuthal angle $\Phi$ should satisfy the following equation: $\Phi=-\varphi_{3,1}$.
Then, the coefficient of cross-correlations $K^-(l=3)$ is given by:
\begin{eqnarray}
K^{-}_{l=3}(\Theta)=\frac{a_{3,0}b\cos\Theta+2|a_{3,1}|c\sin\Theta}{\sqrt{7C(l=3)\left[b^2\cos^2\Theta+2c^2\sin^2\Theta\right]}}\nonumber\\
=\left(\frac{a^2_{3,0}b^2+4c^2|a_{3,1}|^2}{7C(l=3)}\right)^{\frac{1}{2}}\frac{\cos(\Theta-\eta)}{\sqrt{b^2\cos^2\Theta+2c^2\sin^2\Theta}}
\label{cormax}
\end{eqnarray}
where we have used the following definitions: $\varepsilon_{3,0}=b\cos\Theta$ and $|\varepsilon_{3,1}|=c\sin\Theta$ (see Eq.(\ref{vareps2})), $C(l=3)$ is the power of the ILC octupole and 
\begin{eqnarray}
\eta=\tan^{-1}\left(\frac{2|a_{3,1}|c}{a_{3,0}b}\right).
\label{eta}
\end{eqnarray}
\FIGURE{
 \centerline{\includegraphics[scale=0.63]{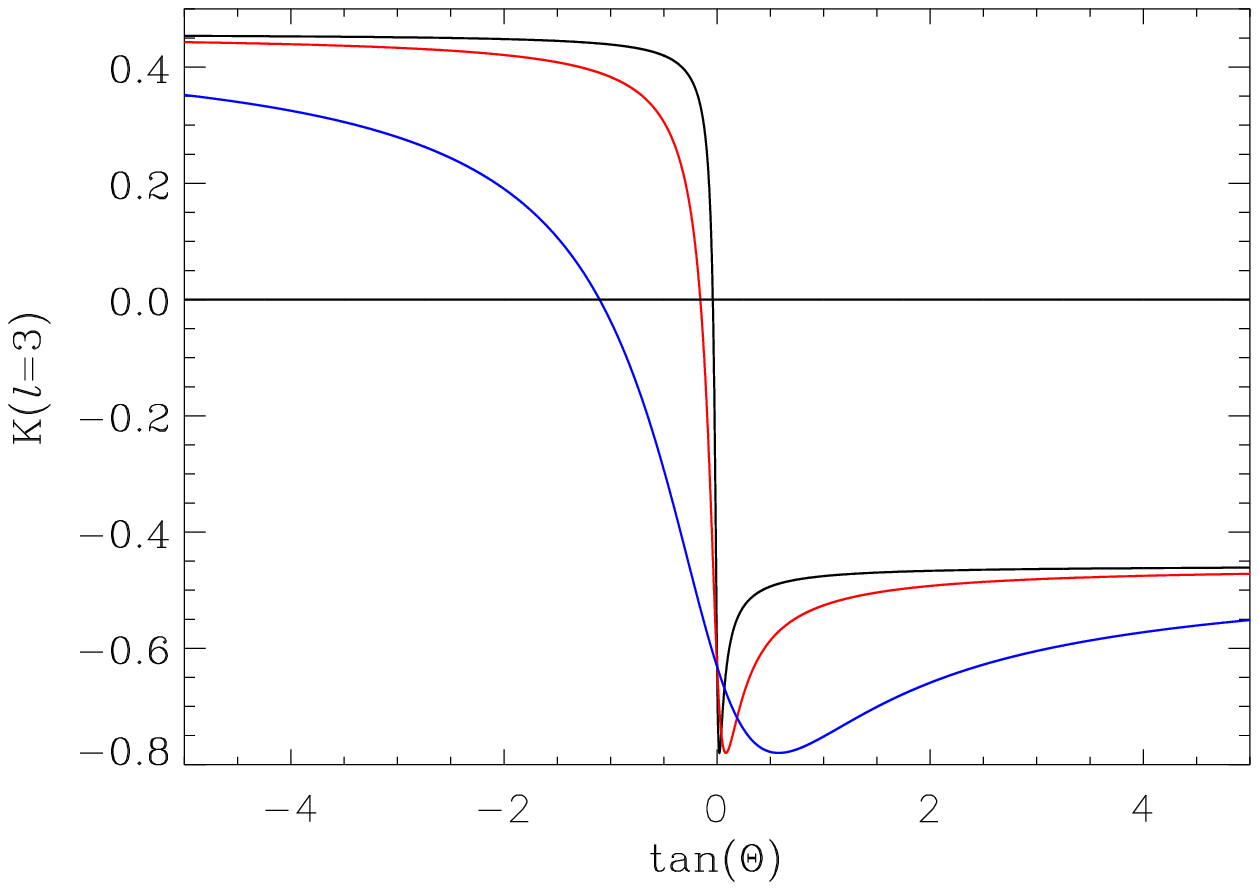}}
\caption{The coefficient $K^{-}_{l=3}(\Theta)$ versus $\tan(\Theta)$ in Ecliptic coordinates. Black line corresponds to $H=15^o$, the red line indicate the model with $H=30^o$, and the blue line corresponds to the $h=70^o$- KBOz model. 
}
\label{tan} }
The nominator in Eq.(\ref{cormax}) has a point of maxima at $\Theta=\eta$, and it vanishes at $\Theta=\eta+\pi/2$. As it follows from Table \ref{alm}, the amplitudes of the octupole in the Ecliptic coordinates are at the same order of magnitude, while the parameters $b$ and $c$ depend on the power spectrum of the KBO angular anisotropy. This is why we will focus on two asymptotics, $|b|\gg |c|$, and $|b|\ll |c|$, in order to investigate the dependency of the cross-correlations on the particular choice of these parameters. In the case $|b|\gg |c|$, all the power of the $\varepsilon_{l,m}$-signal would be concentrated at $m=0$ mode, and 
\begin{eqnarray}
K^{-}_{l=3}(\Theta)\simeq \frac{a_{3,0}}{\sqrt{7C(l=3)}}\simeq -0.65,\hspace{0.3cm}\Theta\ll \frac{\pi}{2},
\end{eqnarray}
In the opposite case, when $|b|\ll |c|$, the coefficient of cross-correlation is given by
\begin{eqnarray}
K^{-}_{l=3}(\Theta)\simeq \frac{\sqrt{2}|a_{3,1}|}{\sqrt{7C(l=3)}}\simeq 0.46,\hspace{0.3cm}\Theta\gg \tan^{-1}\left(\frac{a_{3,0}b}{2c|a_{3,1}|}\right)
\end{eqnarray}
In Fig. \ref{tan} we show the dependency of $K^{-}_{l=3}(\Theta)$ for different KBO models with $H=15^o,30^o, 70^o$. As is seen, for the KBO models with $H=15^o$ and $H=30^o$, the point of maxima of $K^{-}_{l=3}(\Theta)$ is close to $\Theta\simeq \pi$, while for $H=70^o$ it is close to $\eta_{70}\simeq 138^o$ from Eq.(\ref{eta}). Next Fig. \ref{corodd} illustrate the cross-correlation coefficient for KBO $H=15^o$, $H=30^o$ with $\Theta=\pi$.
\FIGURE{
 \centerline{\includegraphics[scale=0.63]{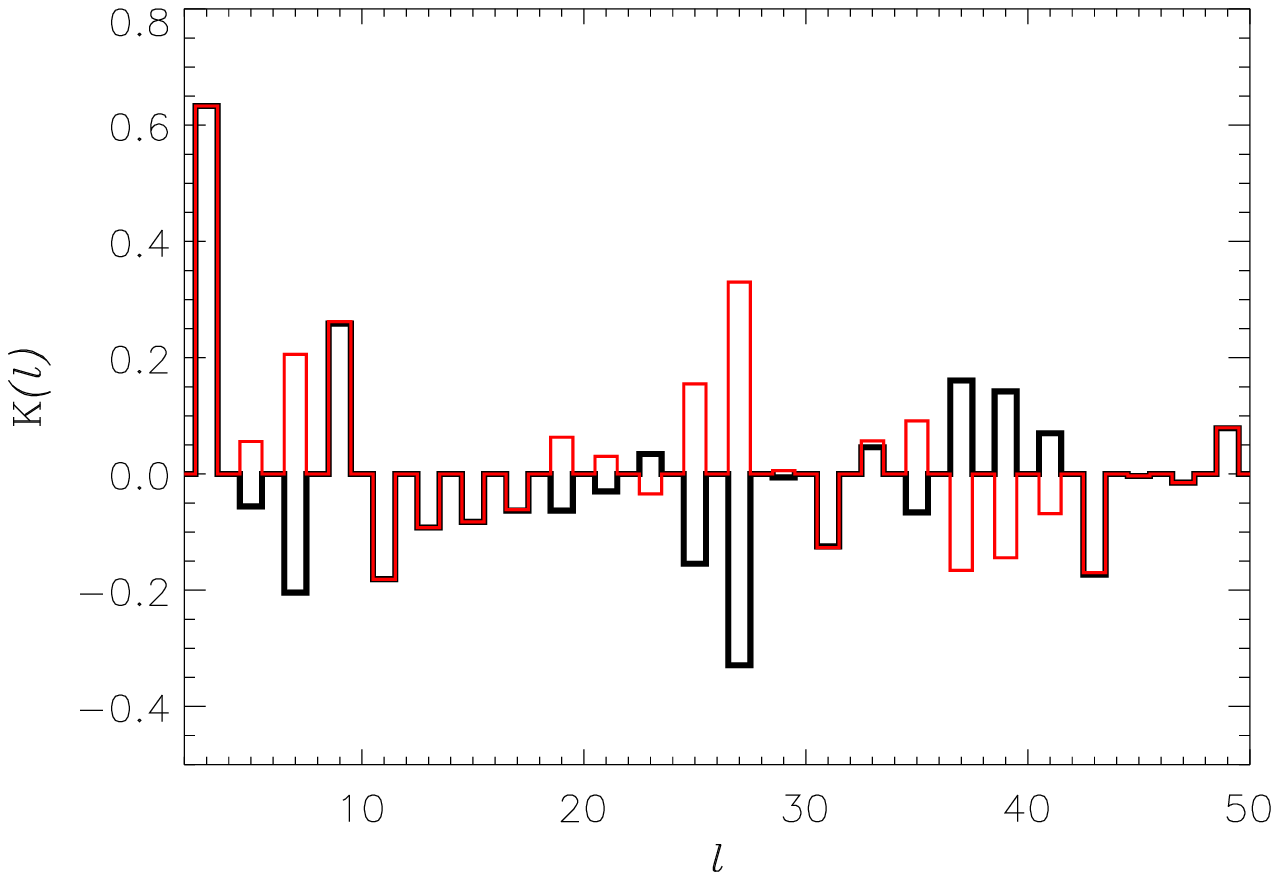}}
\caption{The coefficient $K^{-}_{l}(\Theta)$ versus $l$. Black line corresponds to $H=15^o$ and the red line indicate the model with $H=30^o$.
}
\label{corodd} }
\FIGURE{
 \centerline{\includegraphics[scale=0.27]{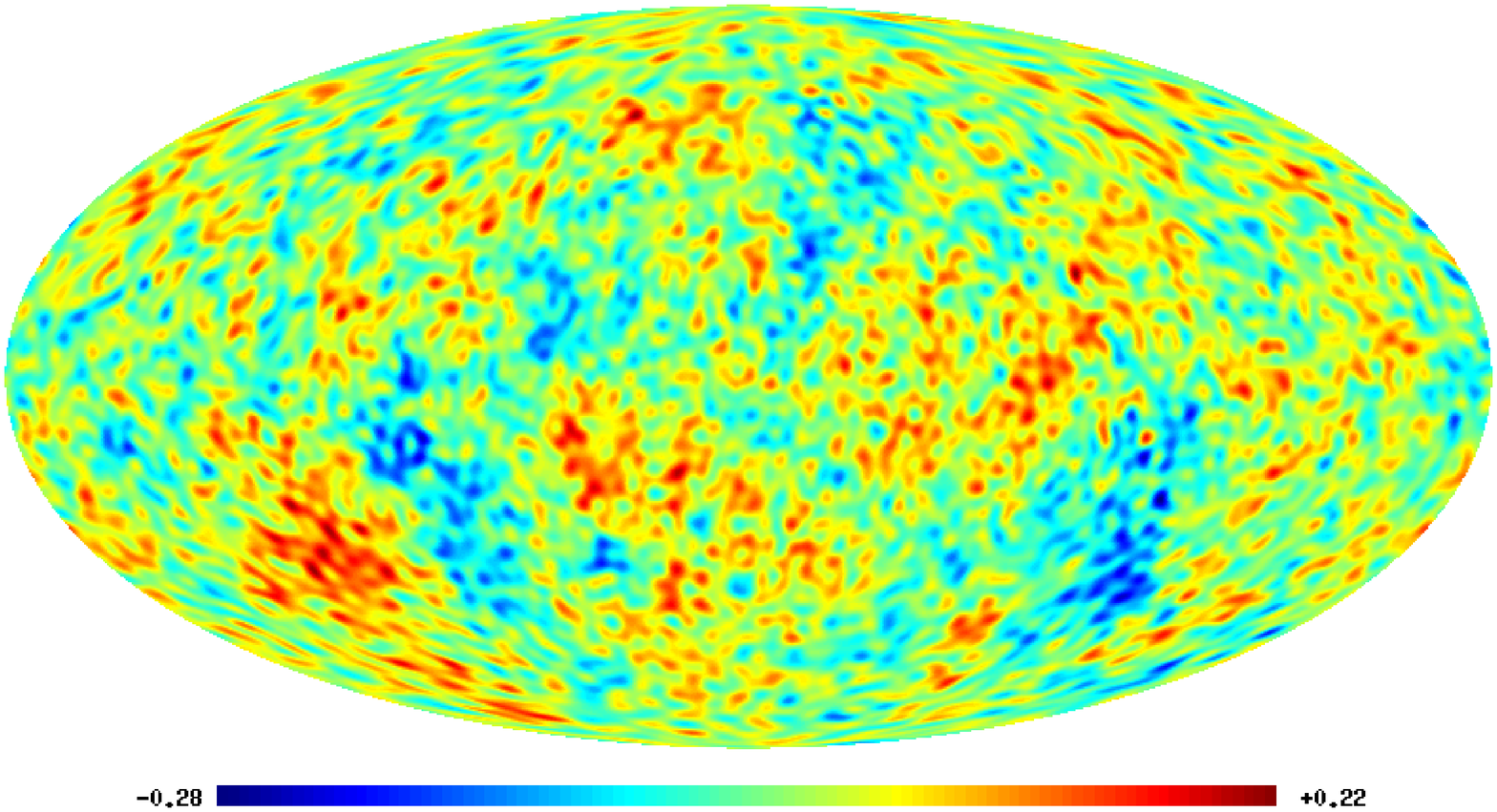}
\includegraphics[scale=0.27]{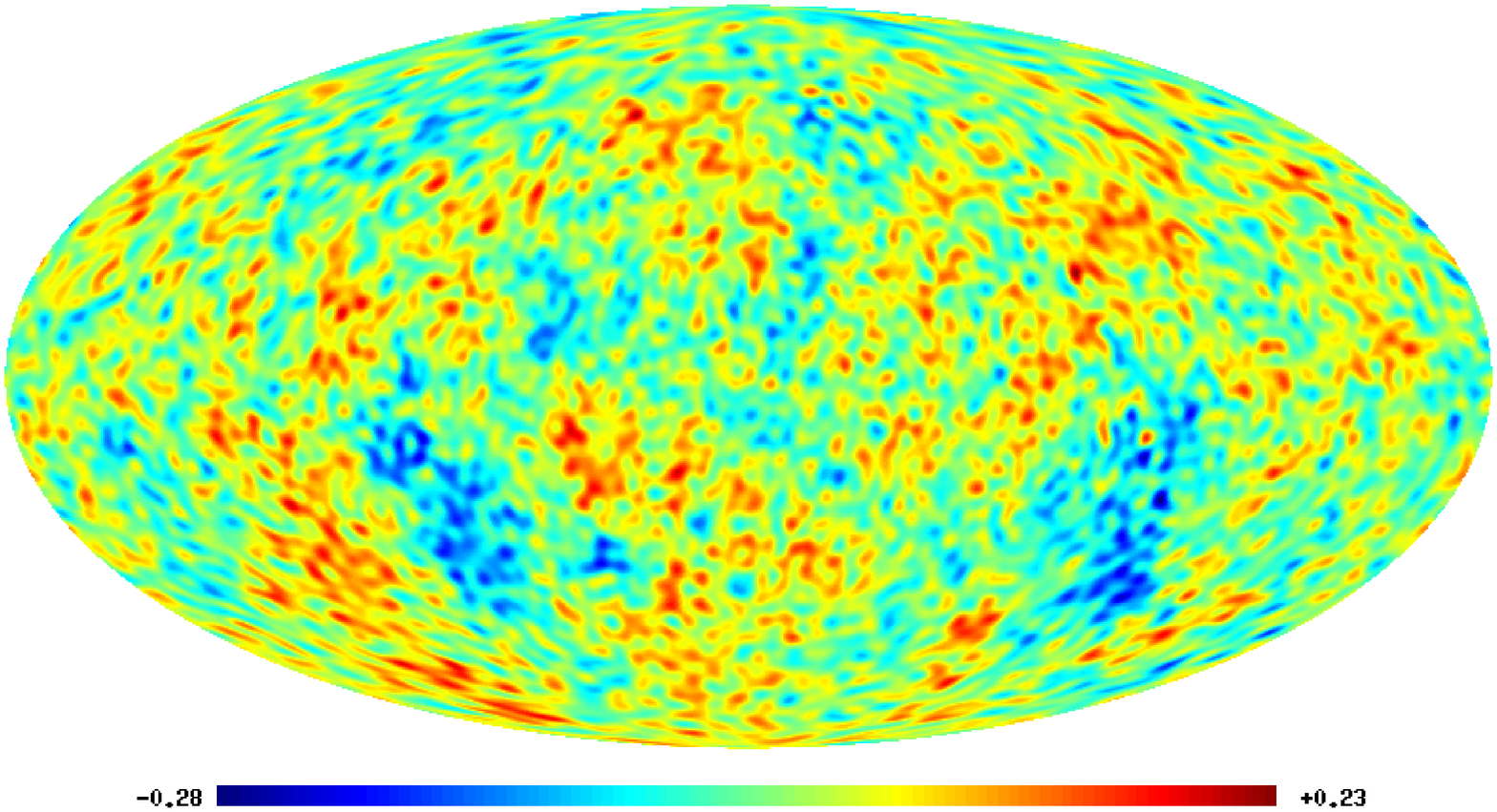}}
 \centerline{\includegraphics[scale=0.27]{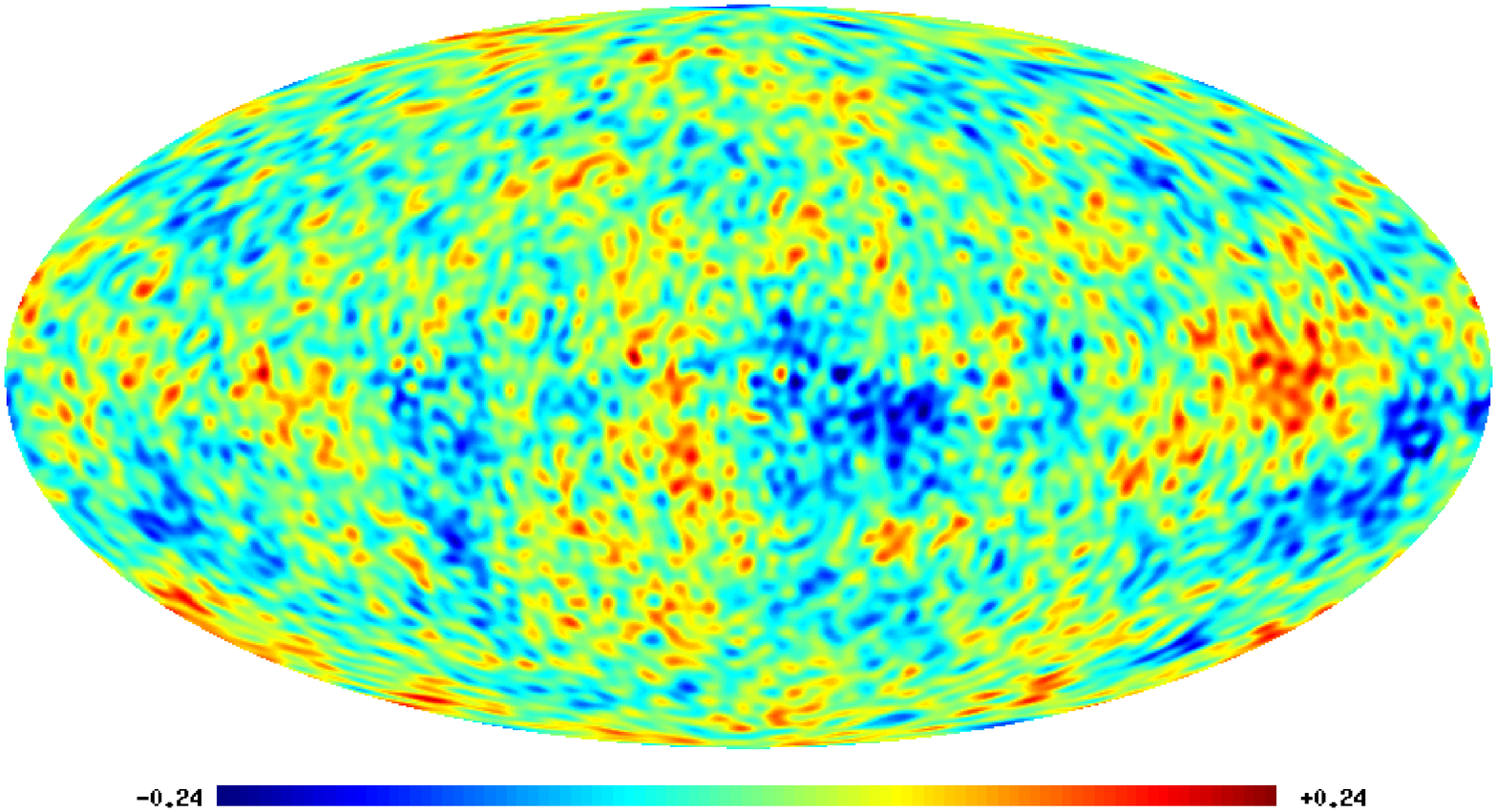}
\includegraphics[scale=0.27]{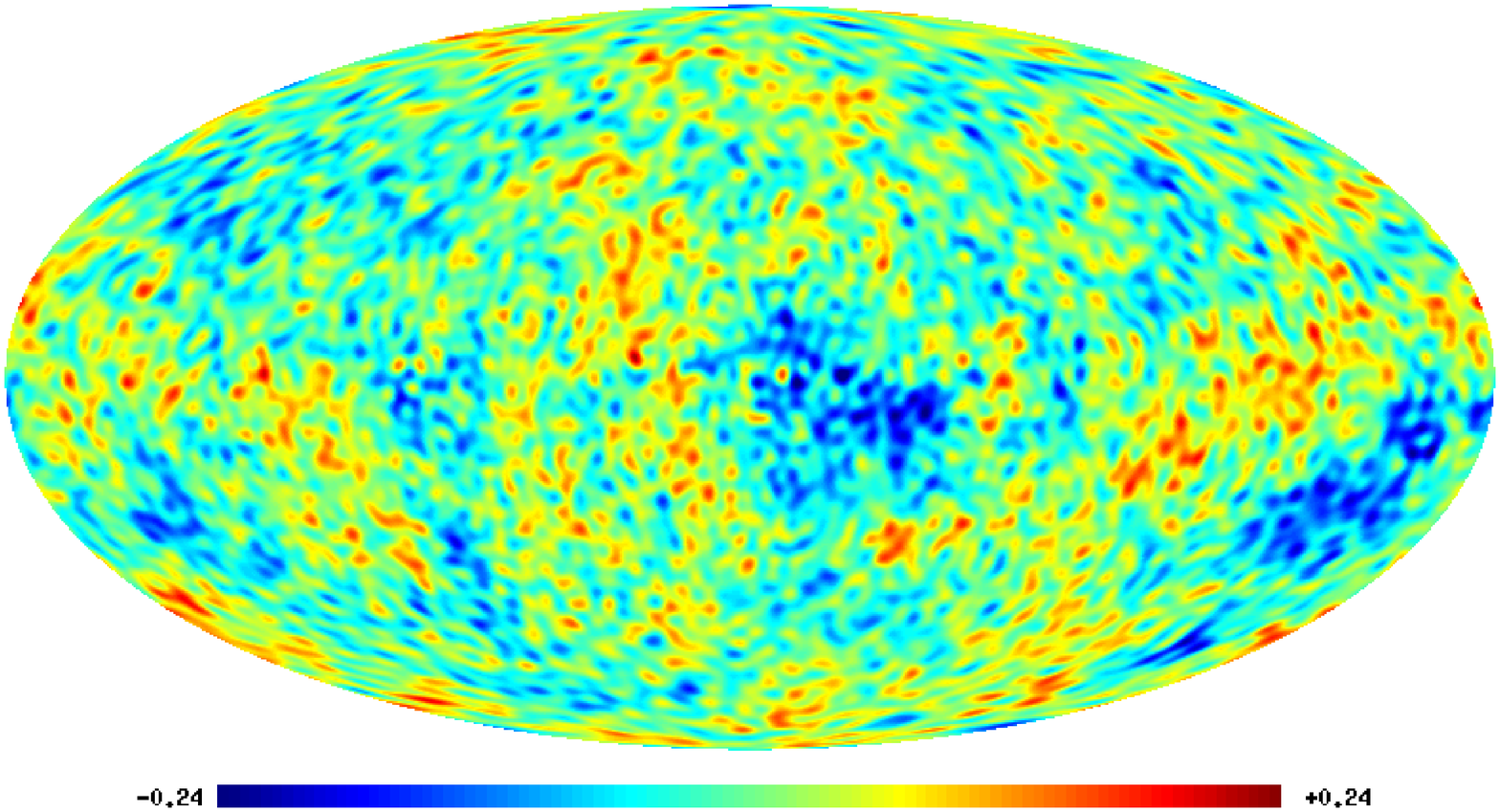}}
 \centerline{\includegraphics[scale=0.27]{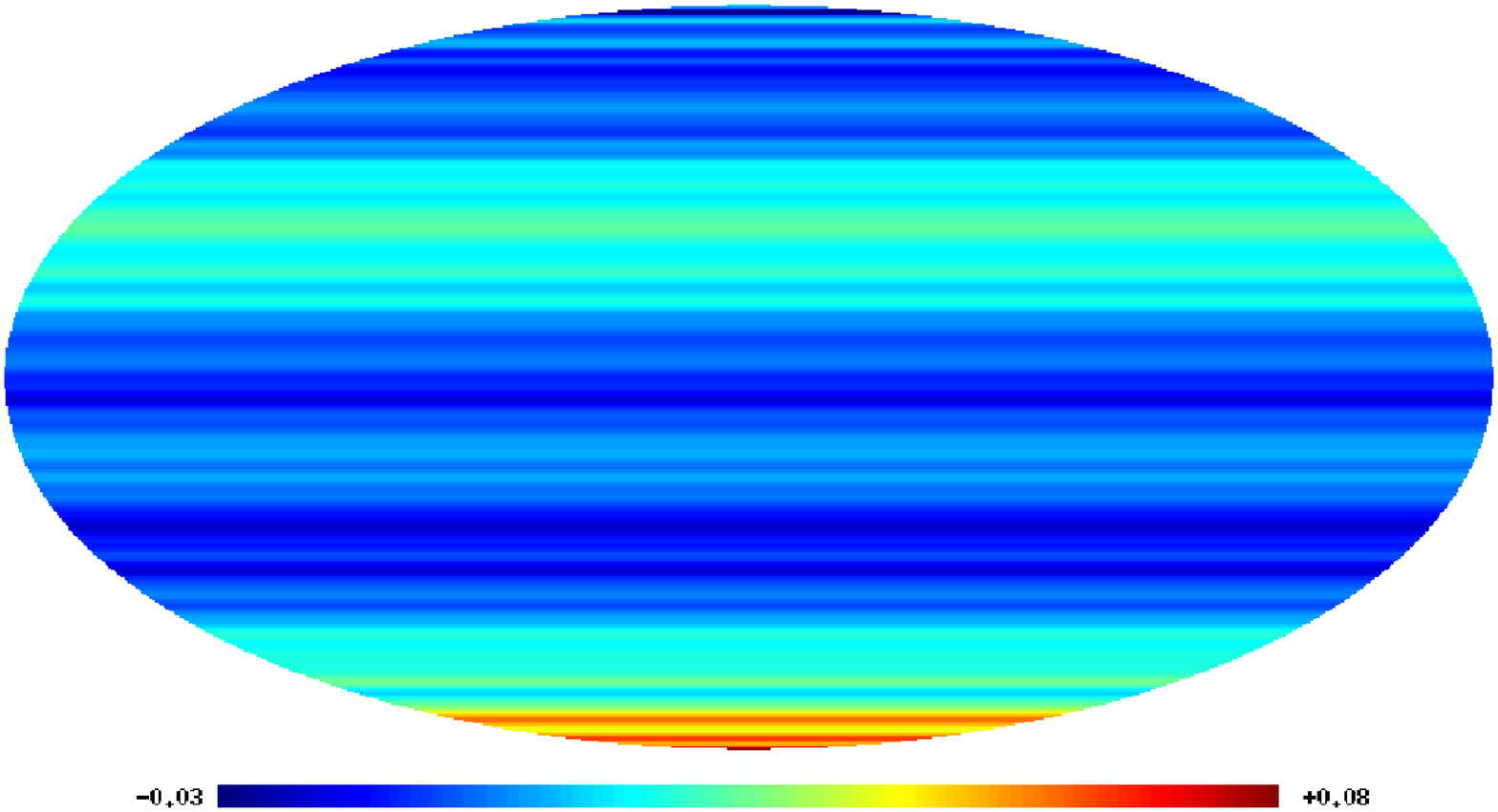}
\includegraphics[scale=0.27]{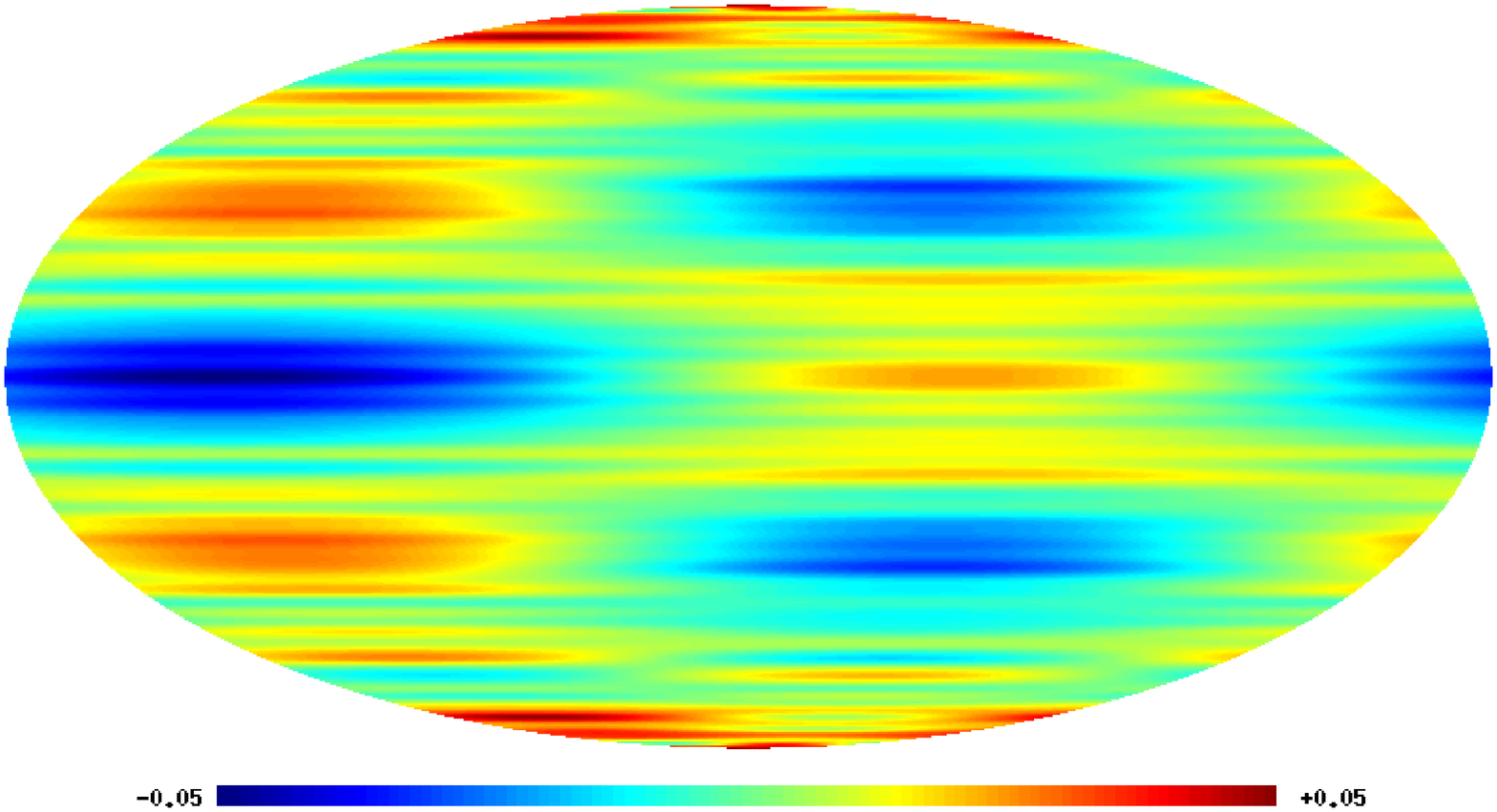}}
 \centerline{\includegraphics[scale=0.27]{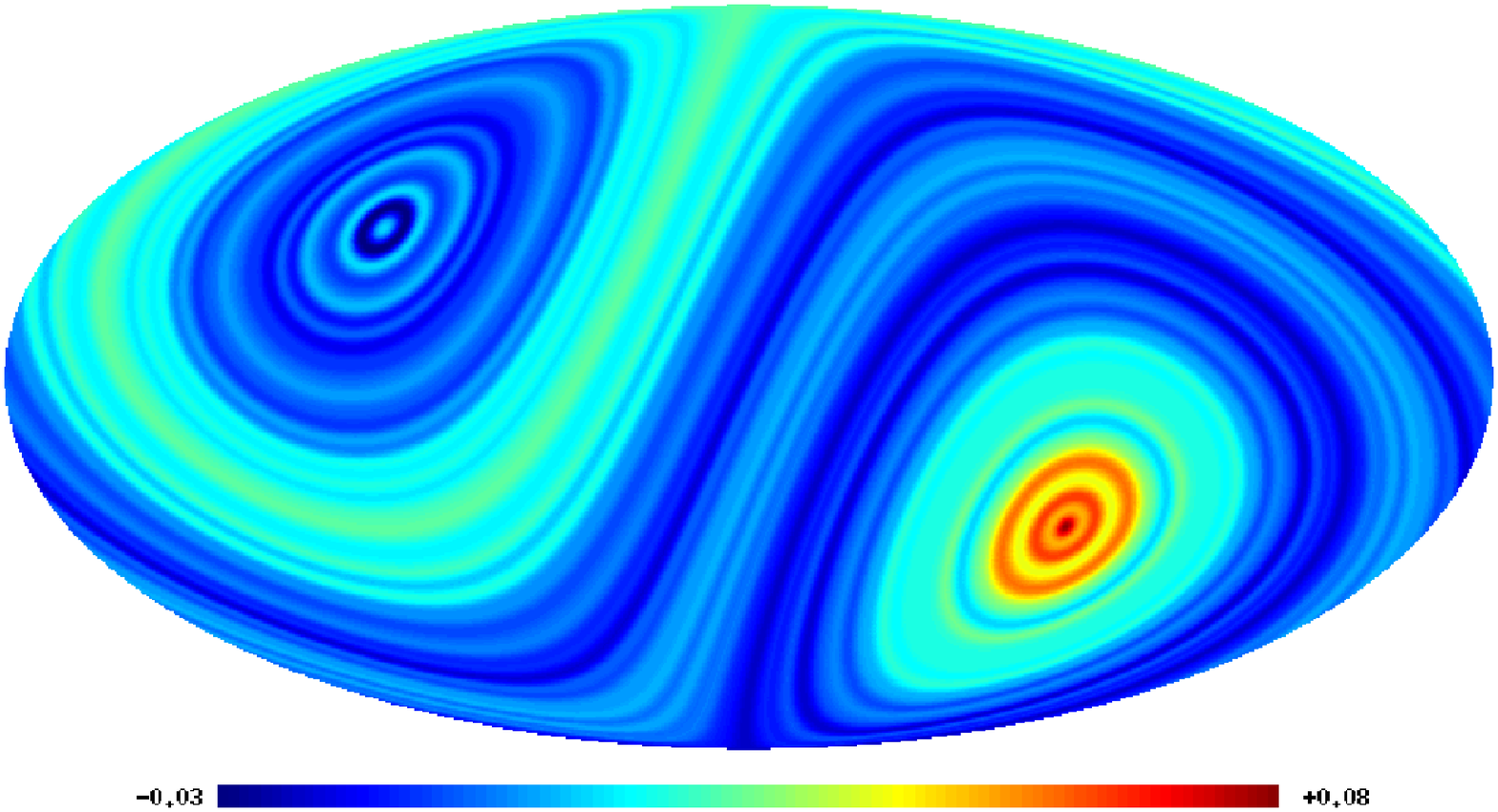}
\includegraphics[scale=0.27]{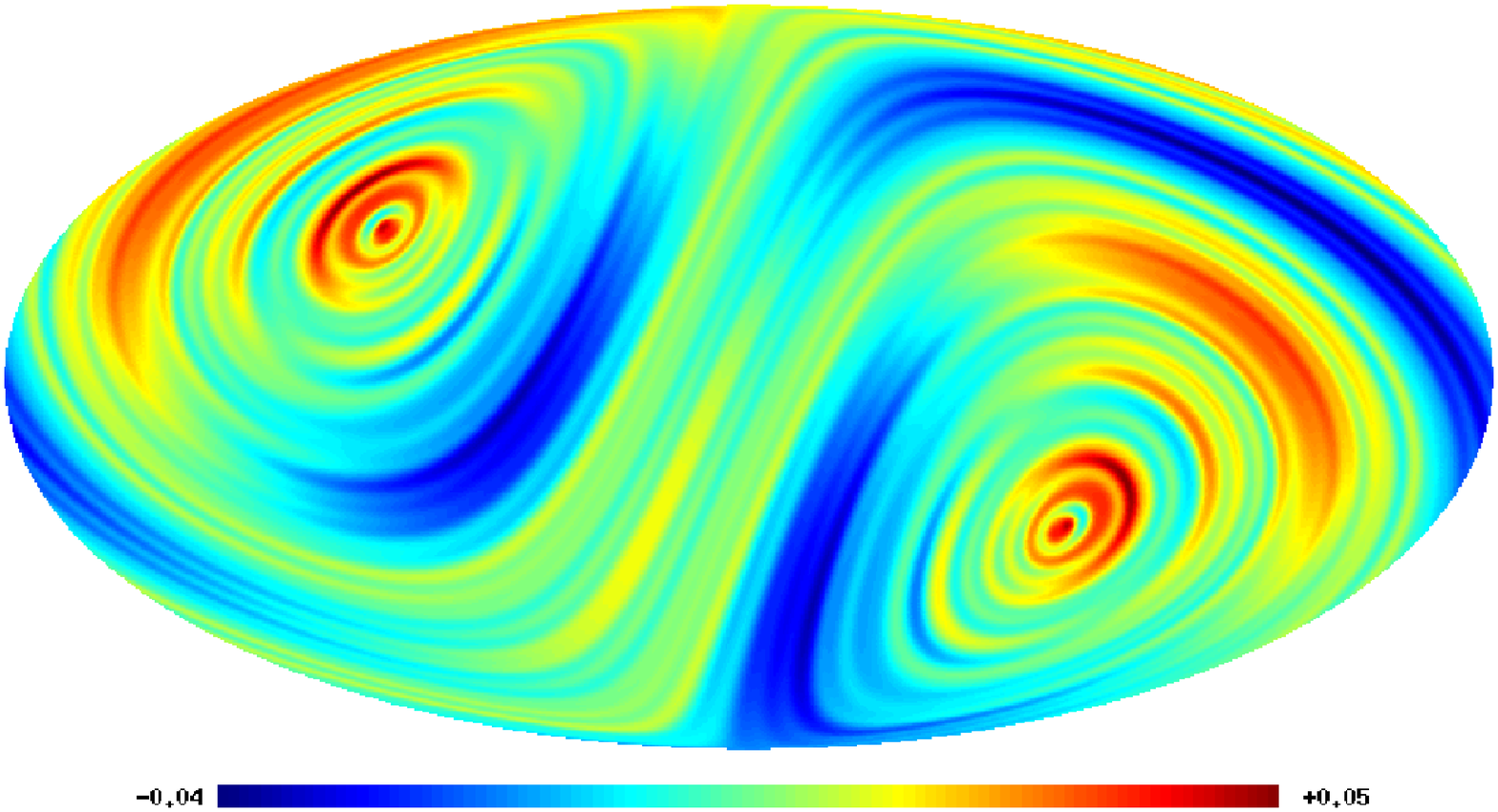}}
\caption{Left column. Reconstructed ILC for the model with $H=30^o$ (in Ecliptic coordinates, top panel). Second from the top is the same signal as the first, but in the Galactic coordinates. The second from the bottom panel, is the map of the dipole modulated KBO foreground (in Ecliptic coordinates). The bottom panel is the same map as above, just in Galactic coordinates. All the maps corresponds to normalization $\Theta=\pi$, $\Phi=-1.227$ rad. Right column. The same as left, but for normalization on the $l=5$ harmonic with $\Theta=\pi/2$, $\Phi=2.2384$ rad.
}
\label{rec} } 
\FIGURE{
 \centerline{\includegraphics[scale=0.27]{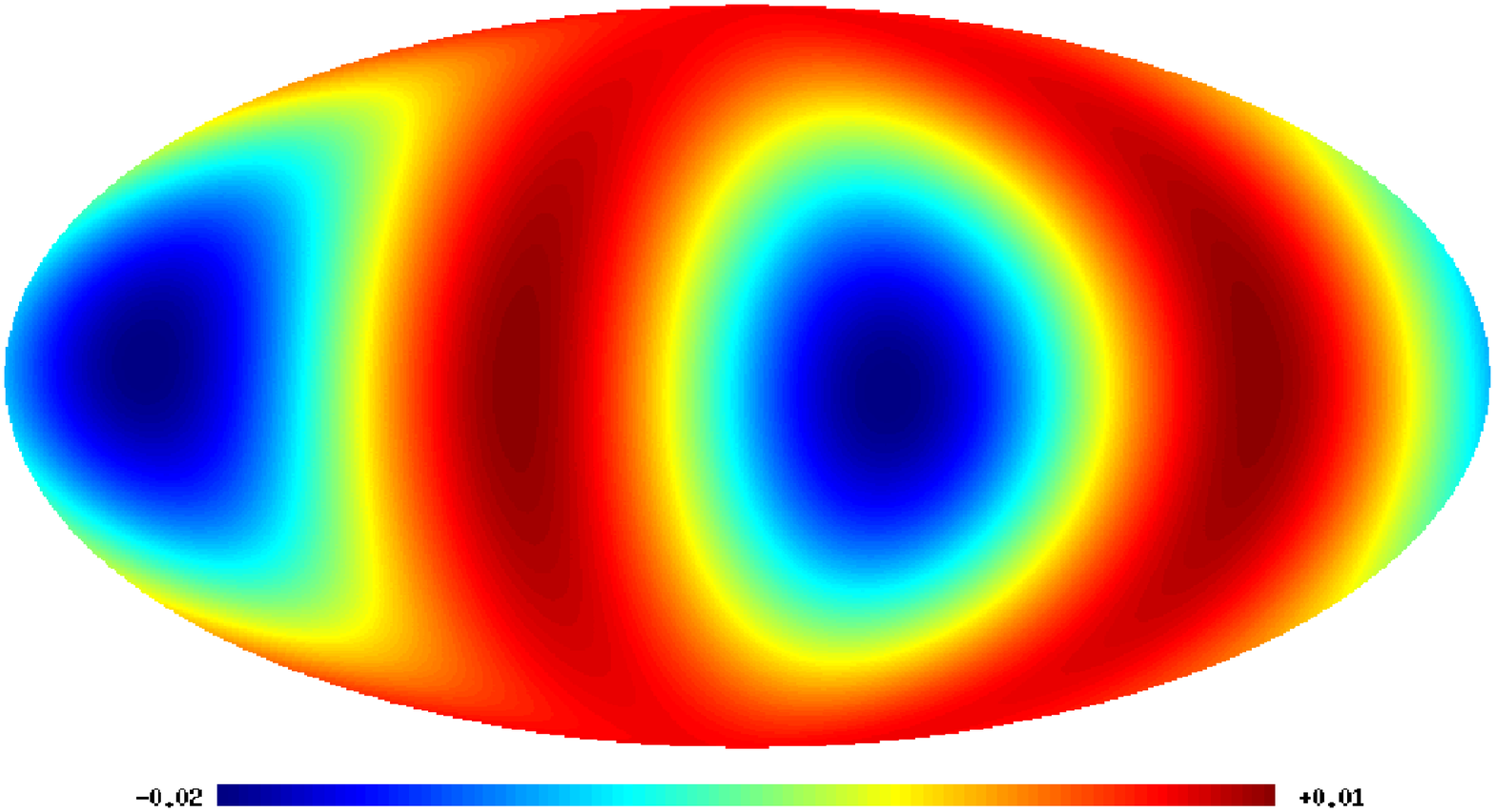}
\includegraphics[scale=0.27]{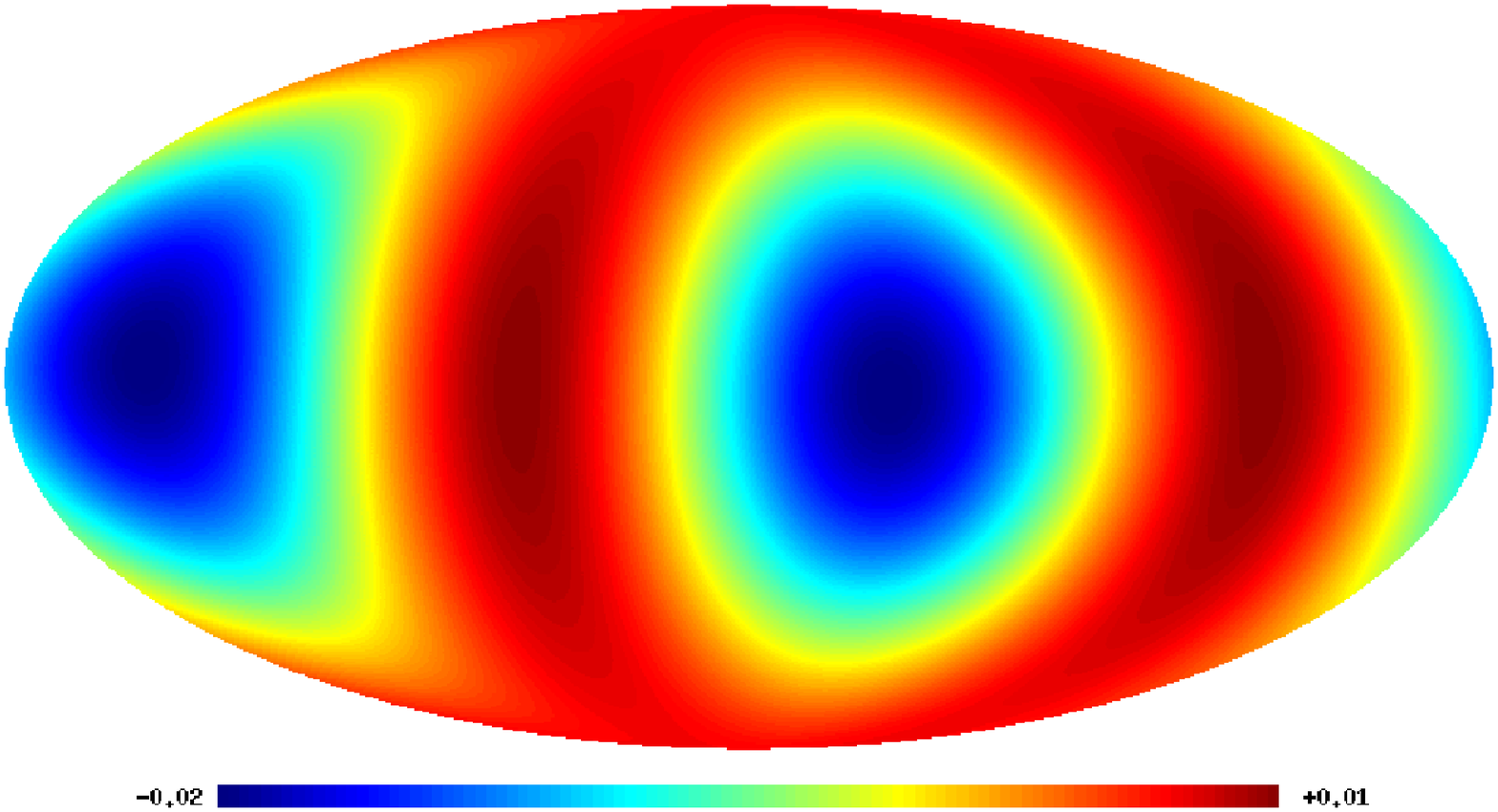}}
 \centerline{\includegraphics[scale=0.27]{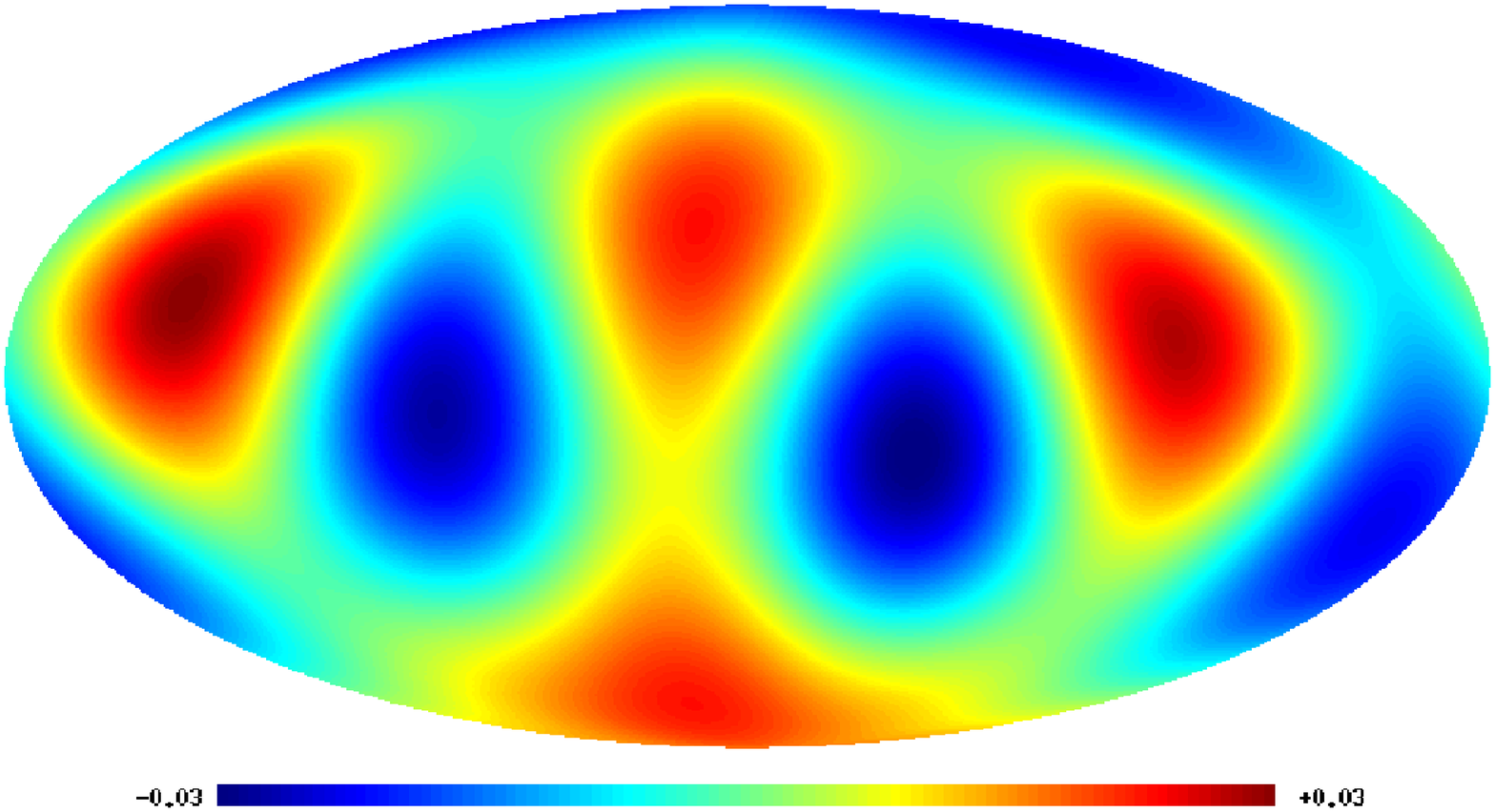}
\includegraphics[scale=0.27]{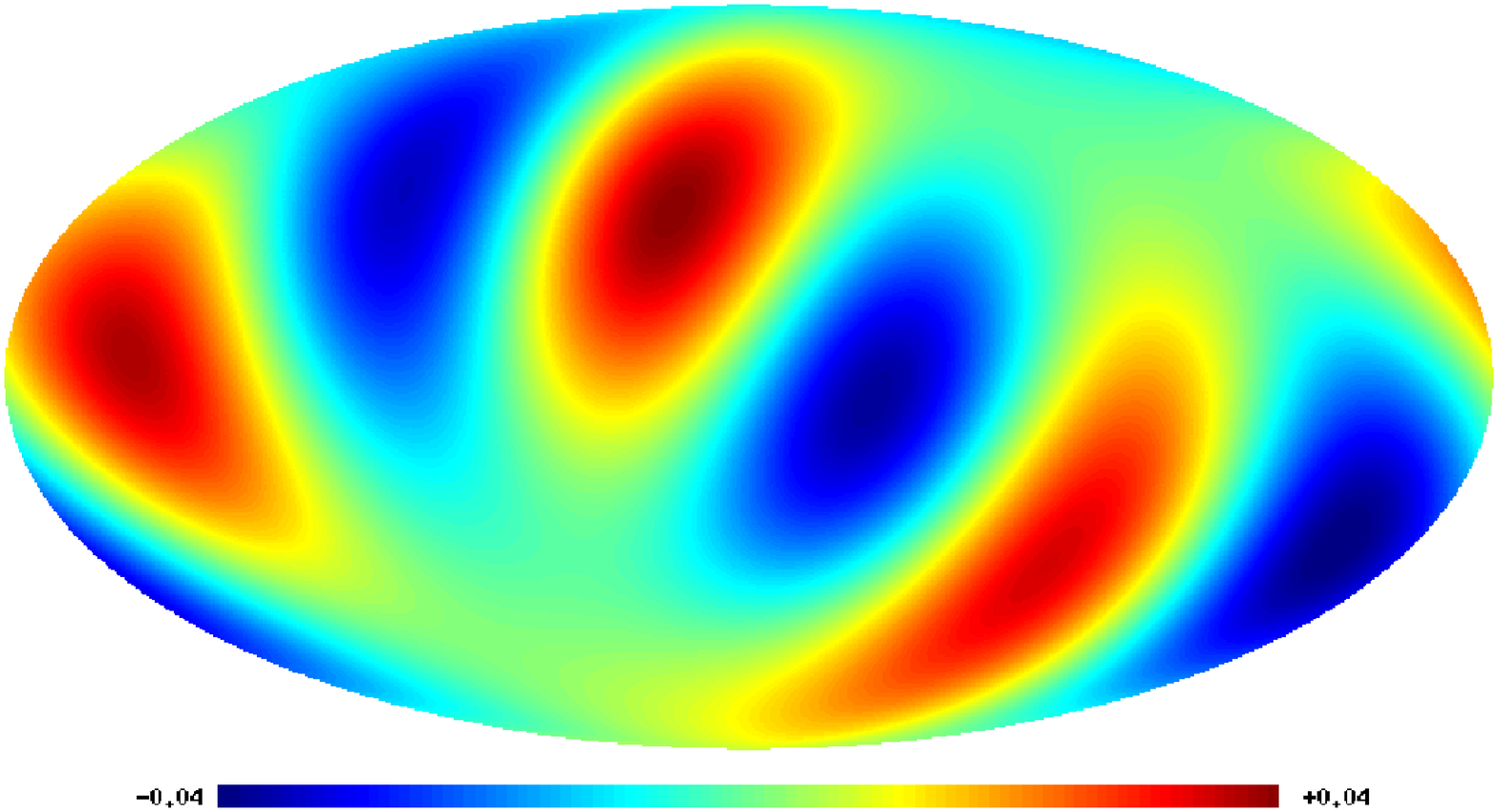}}
 \centerline{\includegraphics[scale=0.27]{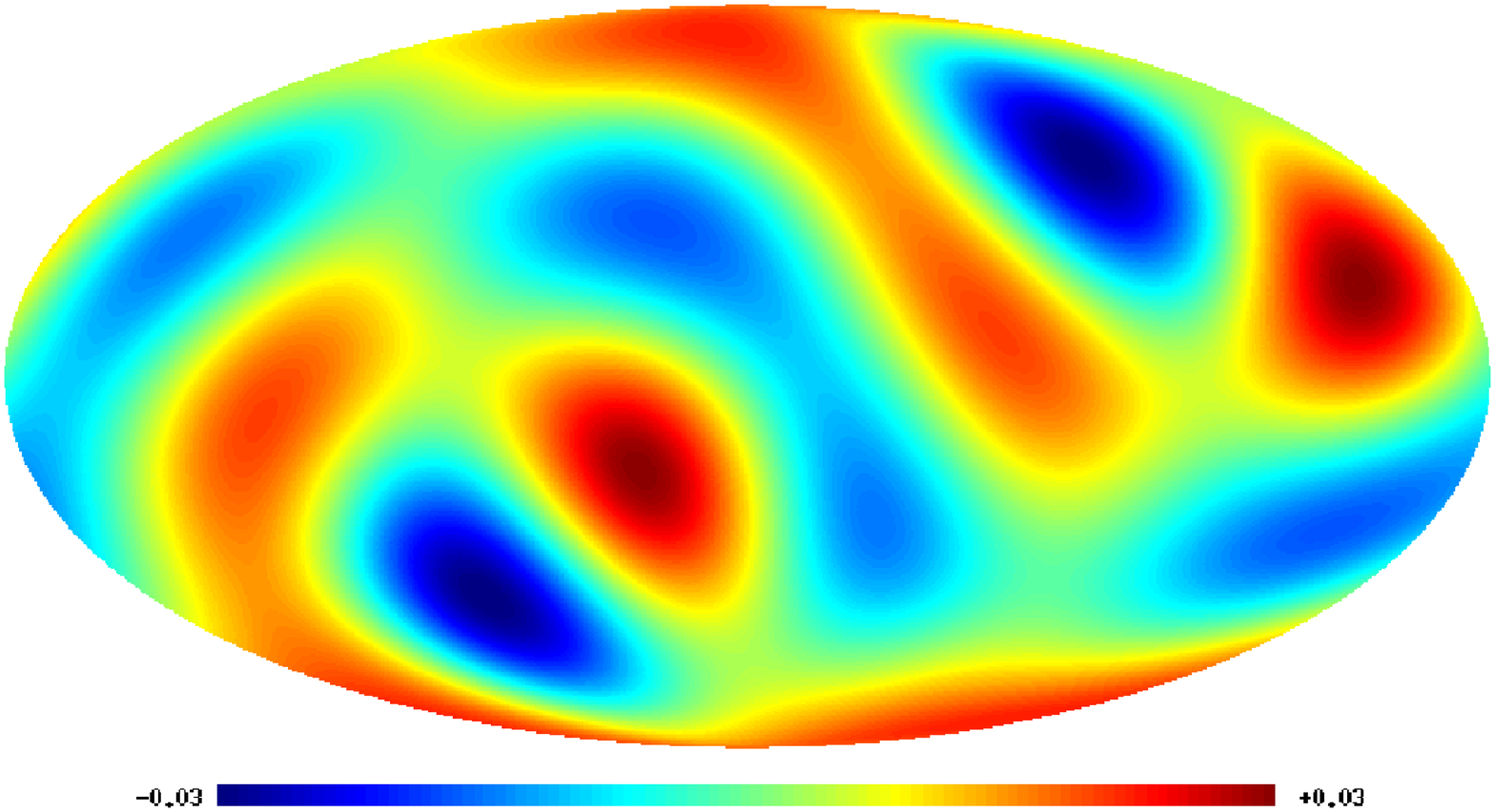}
\includegraphics[scale=0.27]{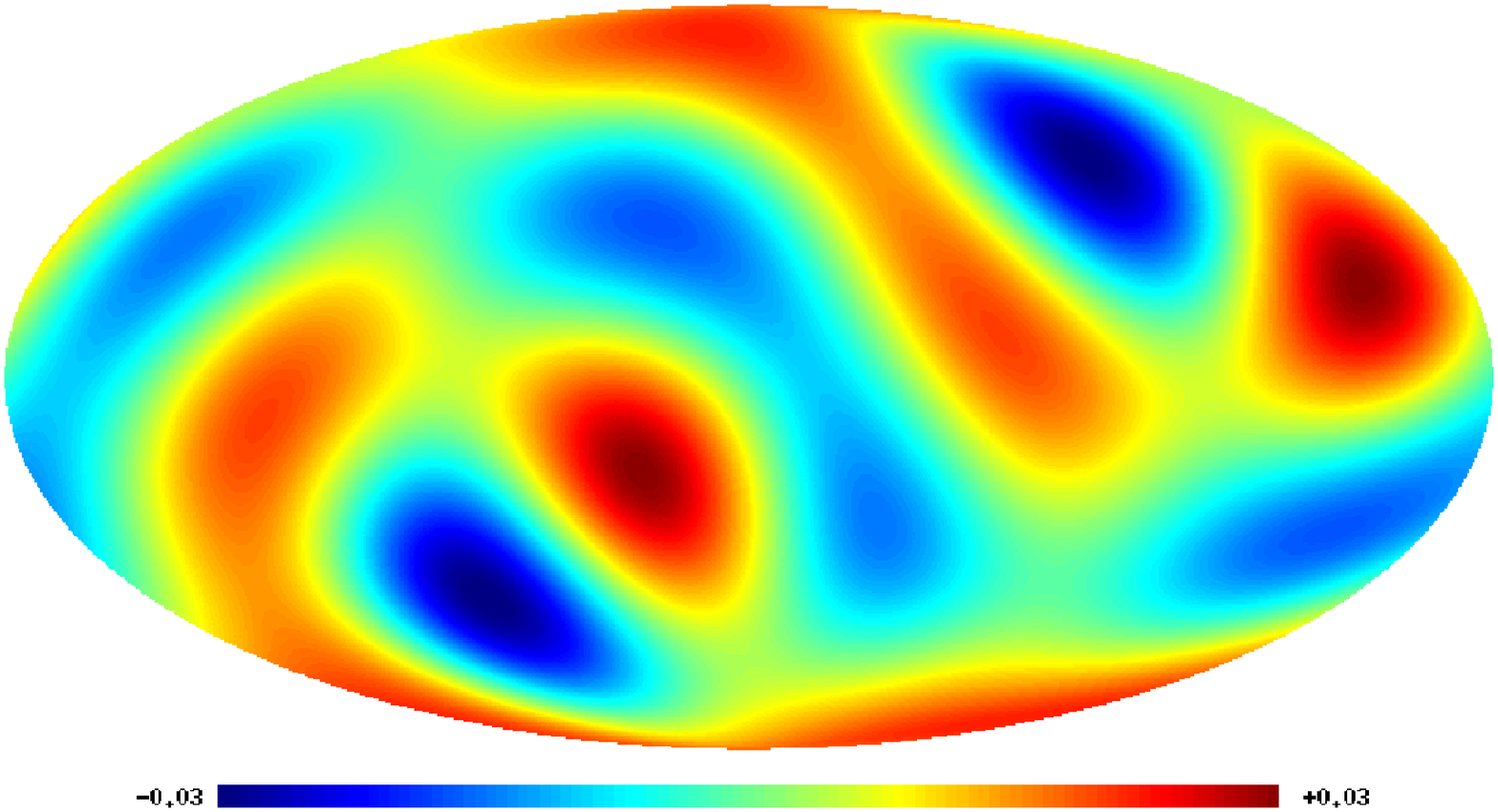}}
 \centerline{\includegraphics[scale=0.27]{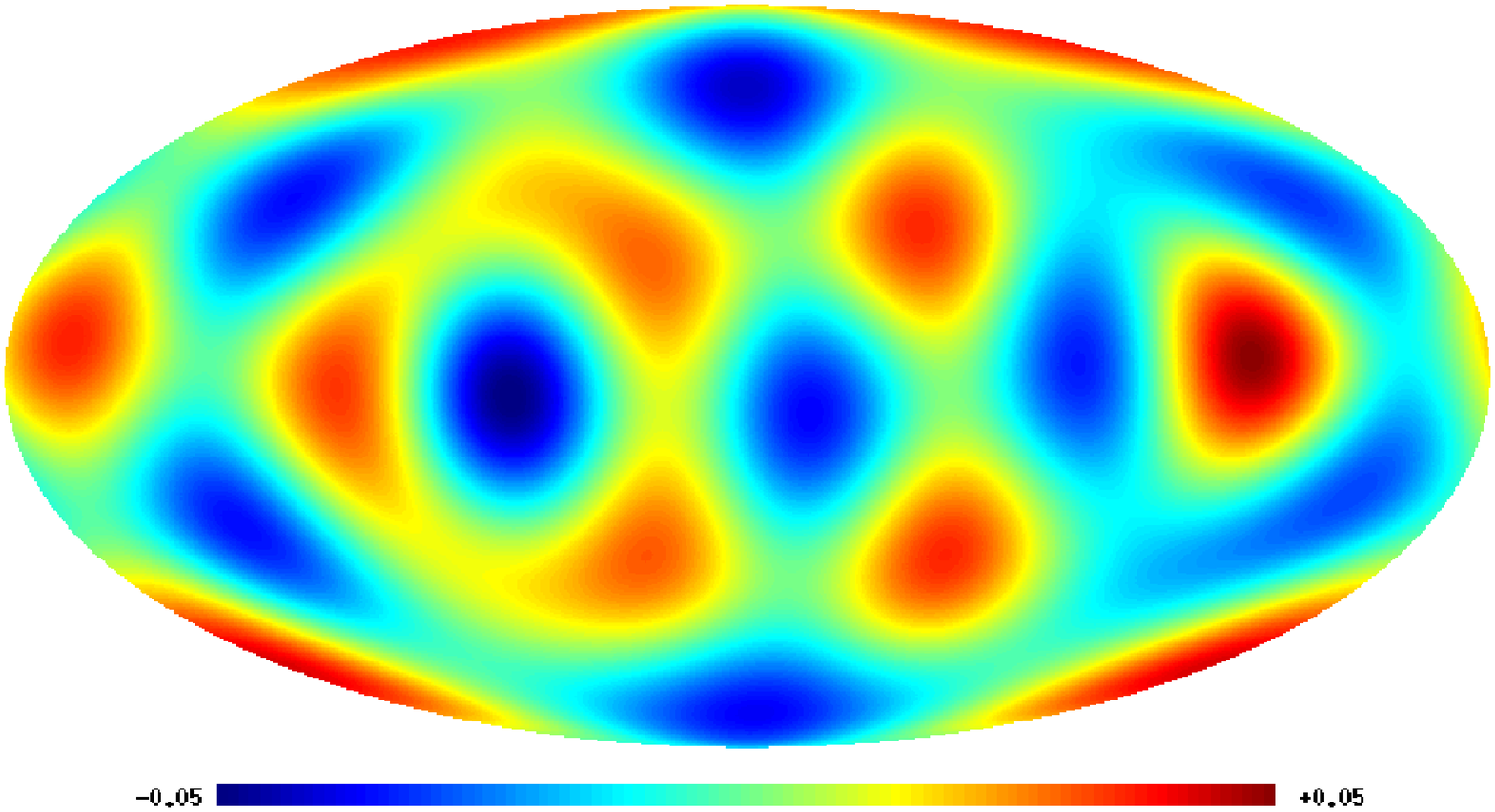}
\includegraphics[scale=0.27]{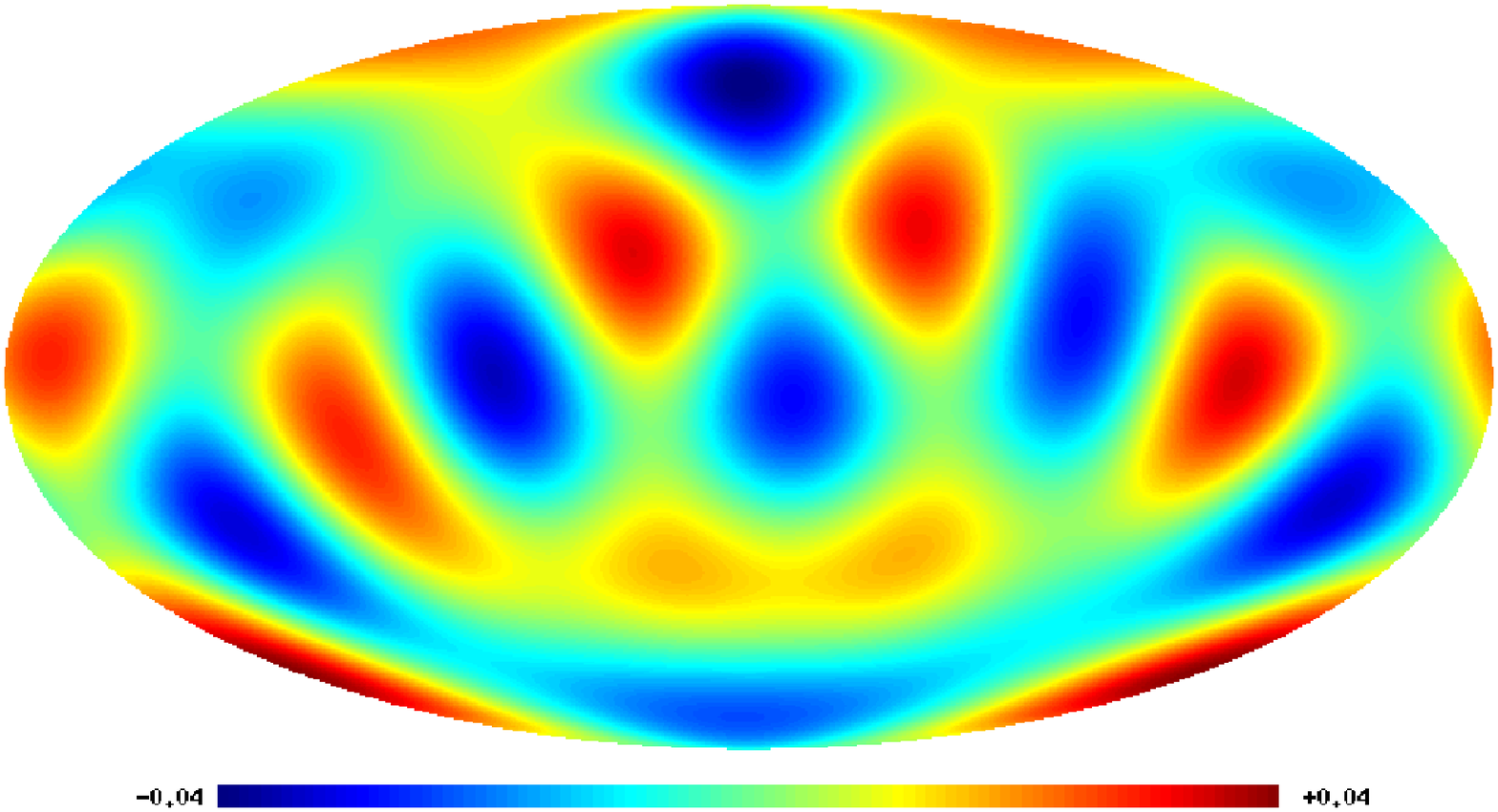}}
\caption{Left column. The map from the top left of Fig. \ref{rec}, but for $l=2,3,4,5$ (from the top to the bottom). Right column. The map from the top right of Fig. \ref{rec}, but for $l=2,3,4,5$ (from the top to the bottom).
}
\label{rec1} }

\subsection{Normalization on $l=5$ component} \label{section:l=5}
As alternative to the normalization of the direction of dipole modulation discussed above, we will in this section use a completely different approach, based on the minimization of power for the $l=5$ multipole in the intrinsic CMB (cf. Table \ref{alm5}). The $l=5$ mode of the power spectrum is one of the major sources of the parity asymmetry, clearly seen in Fig. \ref{powodd} (see next section). 
\TABLE{
\caption{$a_{5,m}$ coefficients of the ILC 7 in the Ecliptic coordinates.}
\begin{tabular}{cccc}
\hline
m &$\Re e(a_{5,m}) $&$\Im m (a_{5,m}) $& $|a_{5,m}|$\\
\hline
$0$& $0.3382 \cdot 10^{-2}$ & 0 & $0.3382 \cdot 10^{-2}$ \\ 
\hline
$1$& $-0.1133 \cdot 10^{-1}$ & $-0.1437 \cdot 10^{-1}$ & $1.1421 \cdot 10^{-2}$ \\
\hline
$2$& $0.1352 \cdot 10^{-1}$ & $-0.1394 \cdot 10^{-1}$ & $1.9419 \cdot 10^{-2}$ \\
\hline
$3$& $-0.1130 \cdot 10^{-1}$ & $0.1008 \cdot 10^{-1}$ & $1.5143 \cdot 10^{-2}$ \\
\hline
$4$& $0.2079 \cdot 10^{-1}$ & $0.2046 \cdot 10^{-1}$ & $2.9169 \cdot 10^{-2}$ \\
\hline
$5$& $0.2692 \cdot 10^{-2}$ & $0.5306 \cdot 10^{-2}$ & $0.5949 \cdot 10^{-2}$ \\
\hline
\end{tabular}
\label{alm5} }
As in the previous section, we will exploit the fact that only the $\varepsilon_{5,0}$ and $\varepsilon_{5,1}$ components of the modulation are non-zero. Taking under consideration, that the phase of the $5,1$ component of the ILC 7 is $ \varphi_{5,1}= -2.23845$, we have adopted $\Phi=-\varphi_{5,1}$ for the azimuthal angle $\Phi$, maximizing the correlations between the $5,1$ component of the ILC and $\varepsilon_{5,1}$. Then, from Eq.(\ref{vareps2}) we get:
\begin{eqnarray}
K^{-}(l=5)(\Theta)=\frac{a_{5,0}\mu\cos\Theta+2|a_{5,1}|\nu\sin\Theta}{\sqrt{11C(l=5)\left[\mu^2\cos^2\Theta+2\nu^2\sin^2\Theta\right]}}
\label{kor5}
\end{eqnarray}
where $\varepsilon_{5,0}=\mu\cos\Theta$ and $|\varepsilon_{5,1}|=\nu\sin\Theta$ (see Eq.(\ref{vareps2}) and $C(l=5)$ is the 
power of ILC for the $l=5$ mode. From Eq.(\ref{kor5}) one finds that the maximum value of $K^{-}(l=5)(\Theta)\simeq 0.28$ can be achieved, if $\Theta\simeq \pi/2$.\\
It should be noted, that the two directions found above ($\Theta=\pi$, $\Phi=-1.227$ rad and $\Theta=\pi/2$, $\Phi=2.2384$ rad) each coincide with objects in the sky. In the case of normalization on $l=3$ ($\Theta=\pi$, $\Phi=-1.227$ rad), the direction of dipole modulation points towards the Carina nebula in the Galactic disc. For the normalization on $l=5$ ($\Theta=\pi/2$, $\Phi=2.2384$ rad), the direction is close to the Coma cluster, at the north Galactic pole. Whether or not this is a coincidence, remains an open question. Also, when looking at the bottom panels of Fig. \ref{rec}, there is a clear similarity between the map of the dipole modulated KBO foreground, and the WMAP map of number of independent observations \footnote{see $http://lambda.gsfc.nasa.gov$}.\\
Having now presented two possible directions for the dipole modulation, we investigate the correlation values between the ILC signal and the KBO-foreground. The optimal conditions for an improvement in the parity asymmetry were given at the end of section \ref{section:Cross_C}. In summary the optimal conditions were a high correlation between the ILC signal and the dipole modulated KBO foreground for odd multipoles, and a low correlation for the even. In fig. \ref{cross_corr} we present the cross correlation (c.f. first part of Eq. \ref{cros}) between the ILC signal and the KBO dipole modulated foreground, for normalization on $l=3$ and $l=5$. The red line symbolizes the limit of cosmic variance \cite{Naselsky1}.
As is evident, the cross correlations does not perfectly fulfill the conditions from section \ref{section:Cross_C} for all the multipoles. It is clear, that there is a relatively strong correlation for several even multipoles, represented by significant peaks. However, at many odd multipoles we also see distinct peaks, with several above the limit of the cosmic variance. The actual influence these correlations will have on the parity asymmetry will be explored in \ref{section:decor}. \\
Notice also the correlation values for $l=3$ in the left panel of fig. \ref{cross_corr}, and $l=5$ in the right. The respective directions of dipole modulation are chosen to maximize these correlation values. One sees further, that the correlation value for the even multipoles is the same in the two panels, showing that the particular direction of dipole modulation only affects the odd multipoles.\\
However a more important point is, that for the entire range of multipoles, from $2$ to $90$, all the correlation values are positive. This is strong evidence, that there is a dipole modulated KBO foreground imbedded in the ILC signal.
\FIGURE{
 \centerline{\includegraphics[scale=0.4]{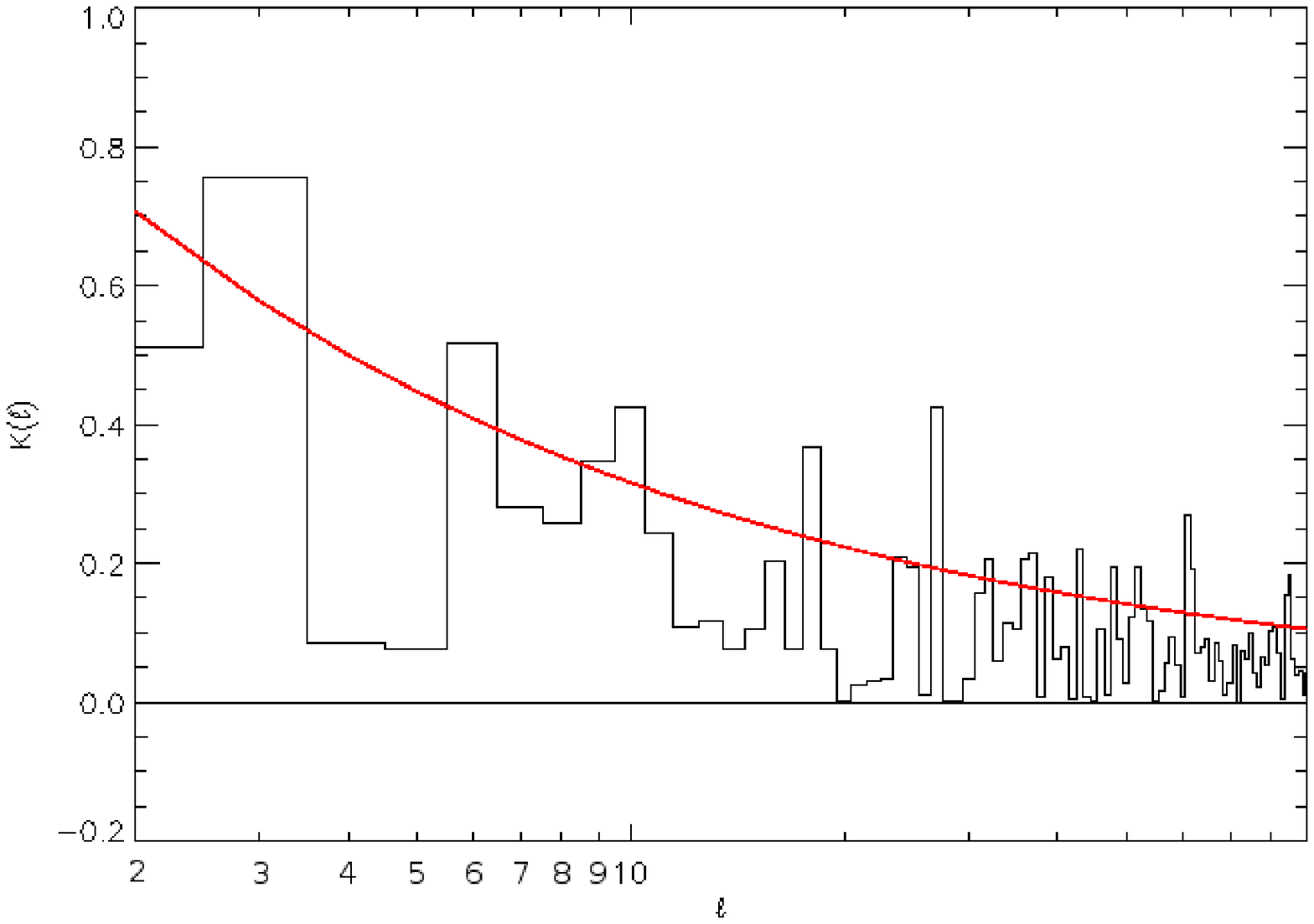}
 \includegraphics[scale=0.4]{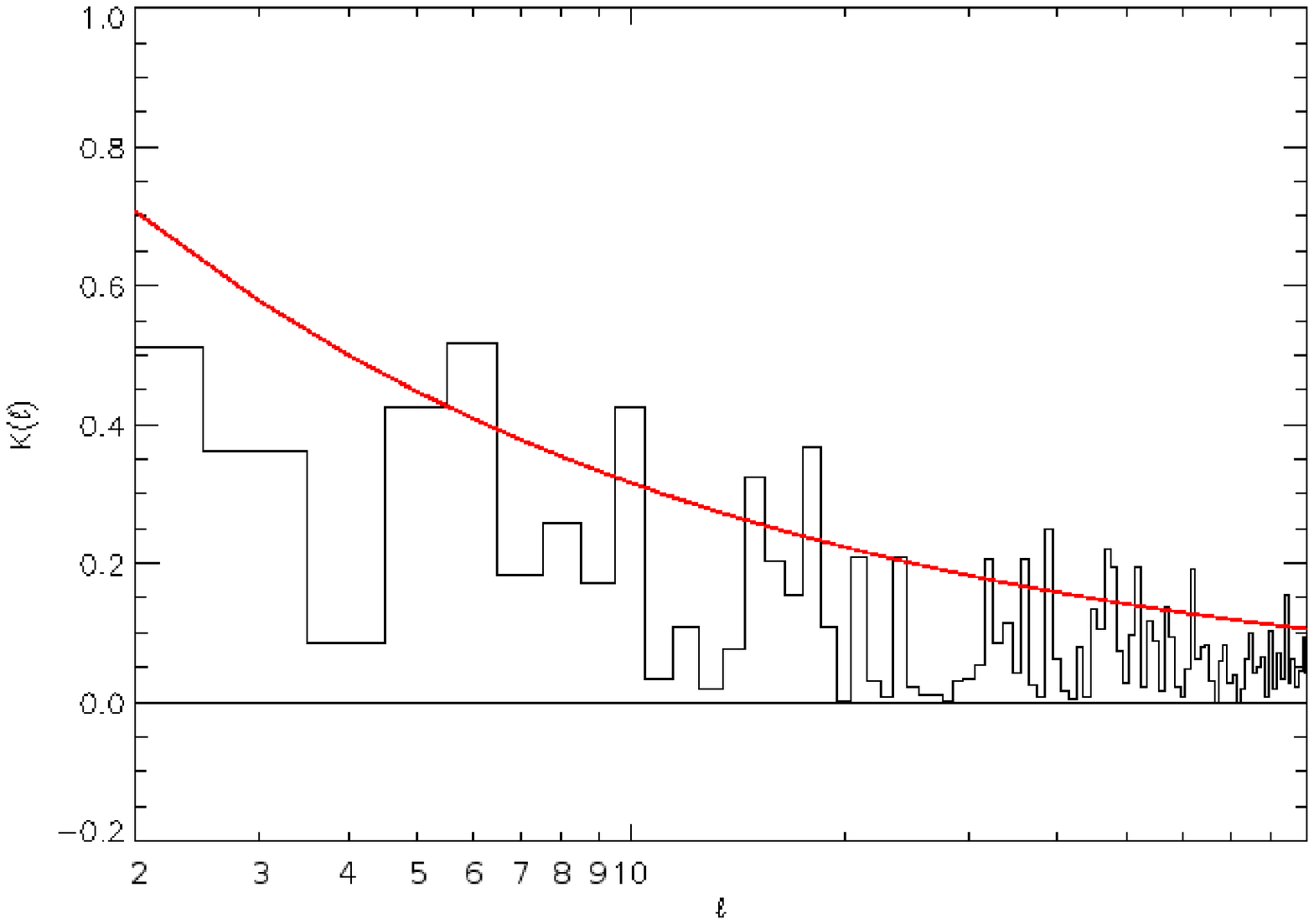}}
 \caption{Left. The cross correlation between the ILC signal and the KBO foreground, with normalization on the $l=3$ multipole in ecliptic coordinates. Right. Same as right, but for a normalization on $l=5$
}
\label{cross_corr} }

\section{De-correlation of ILC and KBO}
In this section we would like to discuss the possible changes in the morphology of the ILC map after removal of the modulations, associated with the KBO emission. We will concentrate on one particular model of the KBO emission, $H=30^o$, using two models of normalization for $\Theta$ and $\Phi$, discussed in the previous section. Hereafter we will call them model-l3 and model-l5. \\
Firstly, we would like to point out that our de-correlation technique is generally unstable, since we cannot recover all the chance-correlations of intrinsic CMB and the KBO, which are sufficient for the low multipole range of the power spectrum (see \cite{Chiang07} for details). This is why de-correlation of the ILC and KBO could only demonstrate the tendency of the changes in the morphology of the ILC, pointing at some general properties of the reconstructed (de-correlated) signal.\\
Secondly, it would be very naive to expect that the simple model of the KBO emission, which is based on a very general symmetry of KBO distribution in space, can significantly improve the ILC map for wide range of multipoles. However, for some particular multipoles even our simple model discussed in Section 2, could be very informative with respect to the possible direction of change in the morphology of the ILC map.\\
Let us discuss this issue in greater detail. Our first step is to assume that the intrinsic CMB, associated with de-correlated ILC map, has zero correlations with the KBO emissivity map, which allows us to set $\kappa(l)=0$ in Eq.(\ref{v}). Then, taking Eq.(\ref{res1}) into account, for the even and odd components independently, we can get the following equation for $\varsigma(l)$ and $\rho(l)$:
\begin{eqnarray}
\varsigma(l) \! = \! \frac{\sum_m \! \left[a_{l,m}\chi^*_{l,m}+a^*_{l,m}\chi_{l,m}\right]}{2\sum_m|\chi_{l,m}|^2} \! = \! K^+(l)\left(\frac{\sum_m|a_{l,m}|^2}{\sum_m|\chi_{l,m}|^2}\right)^{\frac{1}{2}},\nonumber
\end{eqnarray}
\begin{eqnarray}
\rho(l) \! = \!\! \frac{\sum_m \!\! \left[a_{l,m}\varepsilon^*_{l,m} \! + \! a^*_{l,m}\varepsilon_{l,m}\right]}{2\sum_m|\varepsilon_{l,m}|^2} \! = \! K^-(l) \!\! \left(\frac{\sum_m|a_{l,m}|^2}{\sum_m|\varepsilon_{l,m}|^2}\right)^{ \! \! \frac{1}{2}}
\label{decor}
\end{eqnarray}
Then, taking the particular models for $\chi_{l,m}$ and $\varepsilon_{l,m}$, discussed in the previous section, we can estimate the reconstructed signal from ILC 7, shown in Fig. \ref{rec} and Fig. \ref{rec1} for model-l3 and model-l5. In Fig. \ref{powodd} we have plotted the power spectrum of the reconstructed signal with respect to the ILC 7 power.
\FIGURE{
 \centerline{\includegraphics[scale=0.4]{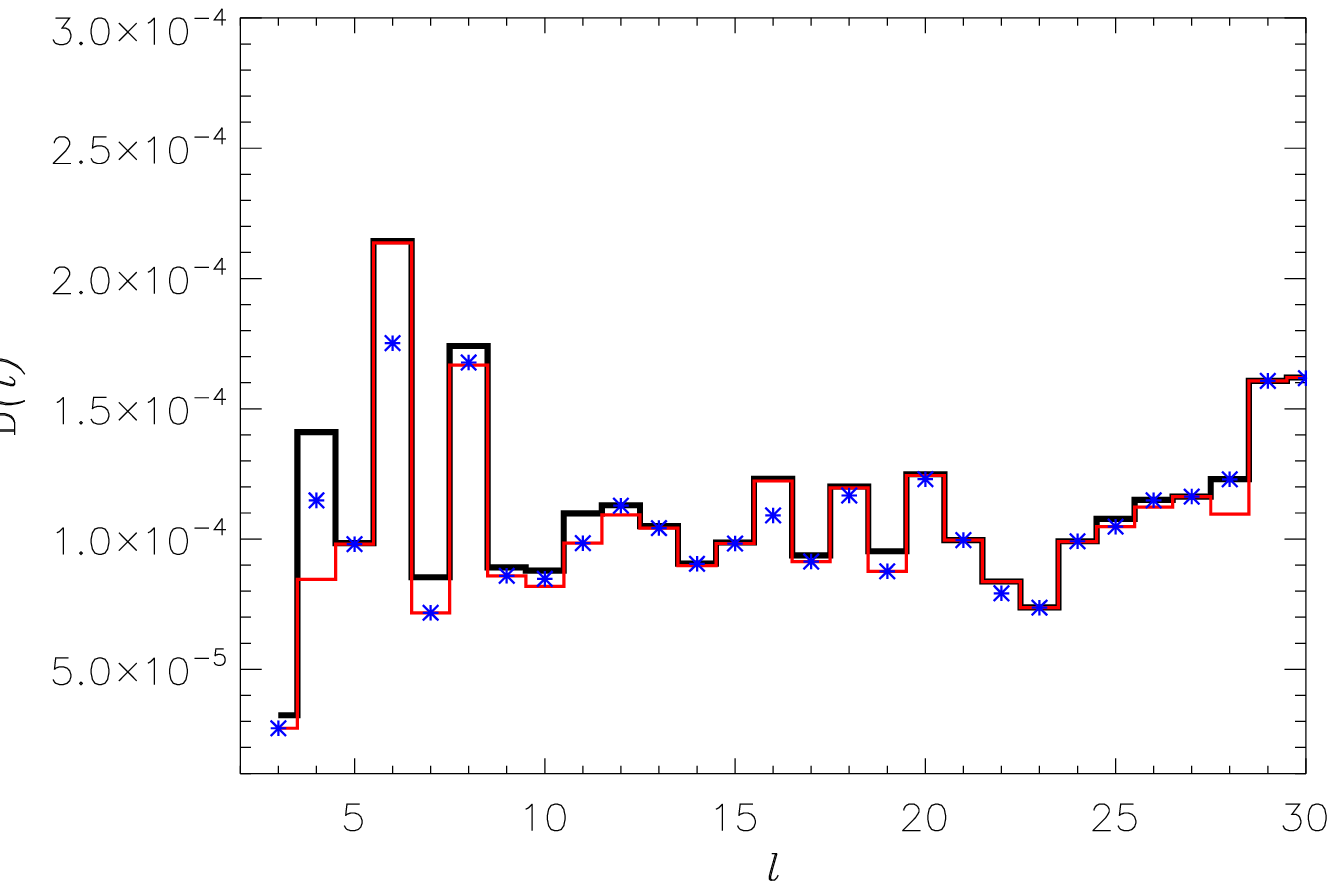}}
\caption{Power spectrum for the reconstructed CMB in the model with $H=30^o$ and normalization of the octupole (the red line), and for normalization of $l=5$ (the blue stars) and for ILC 7 (the black line).
}
\label{powodd} }

\section{Alignment and parity tests for de- correlated CMB}
\label{section:decor}
We now perform a test of the alignment between the quadrupole and octupole for the de-correlated CMB signals, using the publicly available code from ($http://www.phys.cwru.edu/projects/mpvectors/$) for the multipole vector approach from \cite{multipolevectors}.
\TABLE{
\caption{$a_{2,m}$ and $a_{3,m}$ coefficients of de-correlated ILC 7 in Galactic coordinates for the models with normalization $l3$ and $l5$.}
\begin{tabular}{ccccc}
\hline
m &$\Re e^{l3}(a_{2,m}) $&$\Im m^{l3} (a_{2,m}) $& $\Re e^{l5}(a_{2,m}) $&$\Im m^{l5} (a_{2,m}) $ \\
\hline
$0$ & $1.3576 \cdot 10^{-2}$ & 0 & $1.3576 \cdot 10^{-2}$ & 0 \\ 
\hline
$1$& $-1.5904 \cdot 10^{-3}$ & $-1.1121 \cdot 10^{-3}$ & $-1.5904 \cdot 10^{-3}$ & $-1.1121 \cdot 10^{-3}$\\
\hline
$2$& $-7.8456 \cdot 10^{-3}$ & $-1.9363 \cdot 10^{-2}$ & $-7.8456 \cdot 10^{-3}$ & $-1.9363 \cdot 10^{-2}$ \\
\hline
\hline
m &$\Re e^{l3}(a_{3,m}) $&$\Im m^{l3} (a_{3,m}) $& $\Re e^{l5}(a_{3,m}) $&$\Im m^{l5} (a_{3,m}) $ \\
\hline
\hline
$0$ & $-2.3688 \cdot 10^{-2}$ & 0 & $-8.5516 \cdot 10^{-3}$ & 0\\
\hline
$1$& $-1.1804 \cdot 10^{-2}$ & $5.4817 \cdot 10^{-3}$ & $-1.4239 \cdot 10^{-2}$ & $-8.4113 \cdot 10^{-3}$\\
\hline
$2$& $2.3844 \cdot 10^{-3}$ & $5.0355 \cdot 10^{-3}$ & $2.1783 \cdot 10^{-2}$ & $-9.7430 \cdot 10^{-3}$ \\
\hline
$3$& $-1.6411 \cdot 10^{-2}$ & $2.0018 \cdot 10^{-2}$ & $-2.8771 \cdot 10^{-3}$ & $2.6001 \cdot 10^{-2}$ \\
\hline
\end{tabular}
\label{table:2m} }
In Table \ref{table:2m} we present the quadrupole and octupole values for the $l3$ and $l5$ models. Notice that the quadrupoles are the same for the two models. This is because the normalization on $l=3$ and on $l=5$ only affects the odd multipoles differently, while the even are affected in the same way (see for instance Eq. \ref{res}).
Our goal is to test, what the orientation is between the various octupoles from Table \ref{table:2m} and the quadrupole, and in particular the difference between the alignment for pure ILC 7 and for our KBO-foreground cleaned maps.\\
In \cite{multipolevectors}, the authors introduce various approaches for comparing the vectors associated with two different multipoles. We use the statistic of 'oriented area', following the definition from \cite{multipolevectors}, as summarized below.
\begin{eqnarray}
|(\hat{v}^{(l1,i)} \times \hat{v}^{(l1,j)})| \cdot |(\hat{v}^{(l2,k)} \times \hat{v}^{(l2,m)})|
\end{eqnarray}
where $\hat{v}^{(l1,i)}$ and $\hat{v}^{(l1,j)}$ come from the $l1$ multipole and $\hat{v}^{(l2,k)}$ and $\hat{v}^{(l2,m)}$ come from the $l2$ multipole. That is, we cross the $i$'th and $j$'th vector from multipole $l1$, and the $k$'th and $m$'th vector from multipole $l2$, before we finally take the dot product between these two surfaces. Values close to $1$ symbolizes a high level of alignment between the two planes. Generally for a given $l1$ and $l2$ and $i \neq j$ and $k \neq m$, there are $M = l1(l1-1)l2(l2-1)/4$ different products, meaning that for the comparison of quadrupole and octupole, we have three oriented areas ($M1, M2$ and $M3$).\\
In Table \ref{table_mvd}, we have summarized the result of the quadrupole-octupole alignment test for the de-correlated CMB signal, including the standard result for the ILC 7 map for comparison. \\
As is evident from Table \ref{table_mvd}, we have the usual strong alignment between quadrupole and octupole for the ILC. On the other hand, for the KBO-foreground cleaned signal, we see significantly reduced alignments among the three oriented areas. In summary, when we remove the KBO contribution from the ILC signal, to reduce the parity asymmetry, we see a weaker (statistically negligible) correlation between the quadrupole and octupole, than is the case for the ILC 7 map. 
\TABLE{
\caption{Result of the oriented area test between the quadrupole- and octupole, for the ILC case ($a_{l,m}$) and the KBO cleaned signal ($c_{l,m}$), with $l3$ and $l5$ normalization from Section 7.}
\begin{tabular}{| c | c | c | c | c |}
\hline
	& $M1$ & $M2$ & $M3$ & $mean$ \\
\hline
$Oct.-Quadr.(l3)$	& $0.1860 $ & $0.4433 $ & $0.4658 $ & $0.36505 $ \\
\hline 
$Oct.-Quadr.(l5)$	& $0.41567 $ & $0.32210 $ & $0.3879 $ & $ 0.3752 $ \\ 
\hline
$Oct.-Quadr.(ilc)$	& $0.7478 $ & $0.7037 $ & $0.7569 $ & $ 0.7361 $ \\ 
\hline
\end{tabular}
\label{table_mvd} }
\FIGURE{
 \centerline{\includegraphics[scale=0.48]{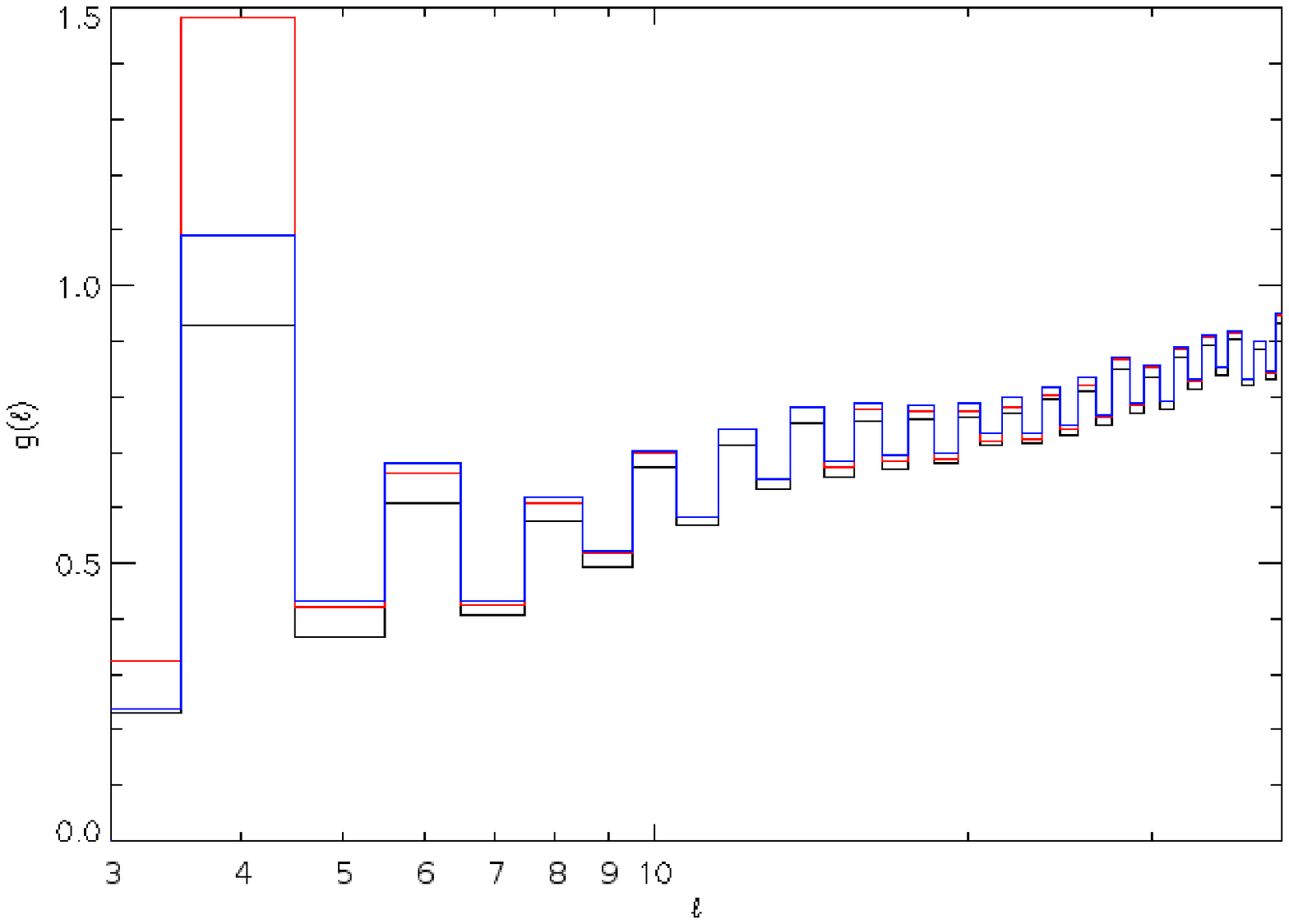}}
\caption{The parity parameter $g(l)$. Black line is for the pure ILC case, red line is for the $l3$ model, and blue line is for the $l5$ model.
}
\label{fig:parity} }
\FIGURE{
 \centerline{\includegraphics[scale=0.48]{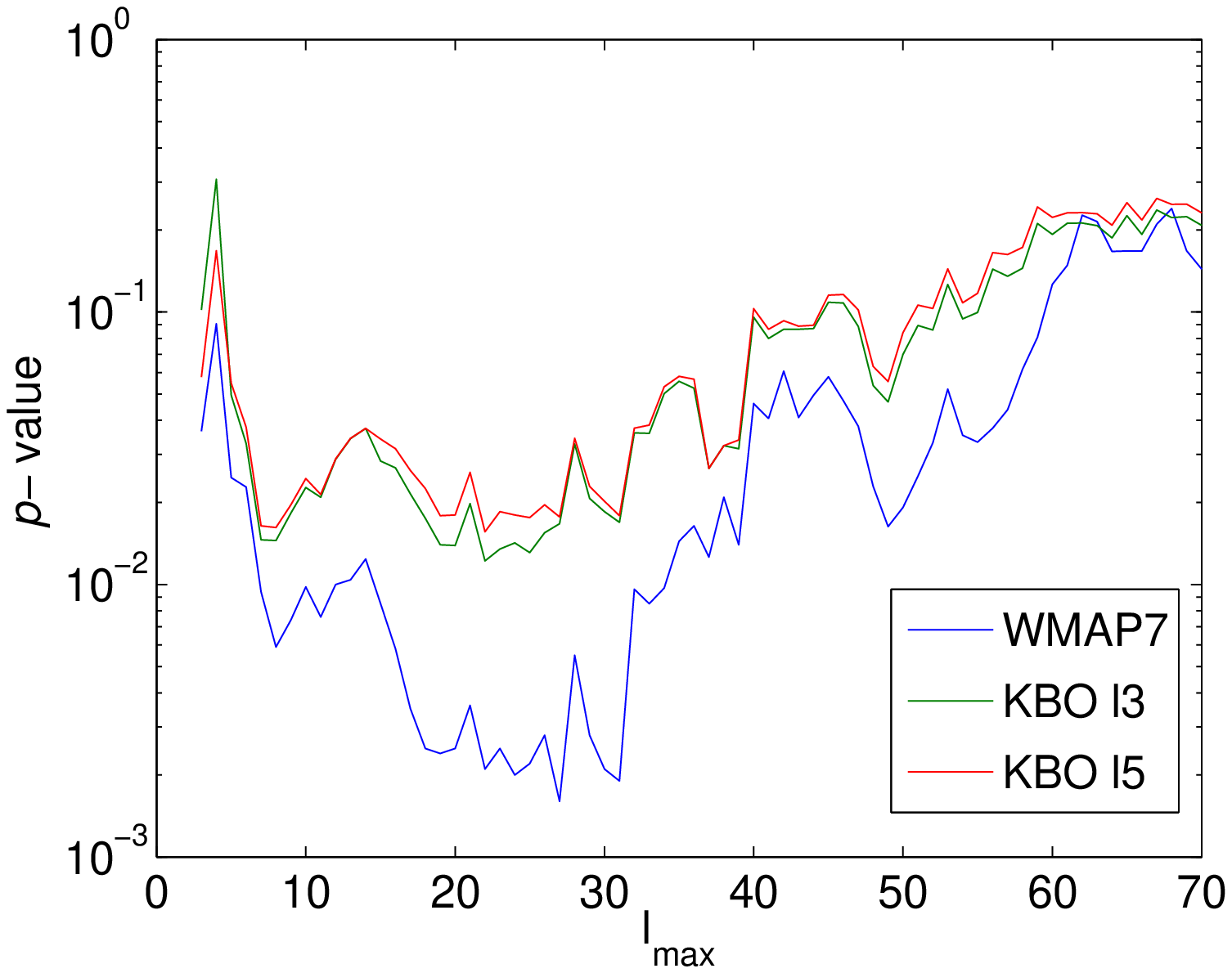}}
\caption{P-value for reconstructed CMB for the model with $H=30^o$ and normalization on the octupole $l3$(the green line), for normalization on $l=5$ (the red line) and for ILC 7 (the blue line).
}
\label{parodd} }
Further, for these two models ($l3$ and $l5$) as well as for the pure ILC case, we have calculated the parity asymmetry parameter $g(l)$ from Eq. \ref{par_est}, with $n_{min}=2$. The results are presented in fig. \ref{fig:parity}. Primarily we see, that the parity parameter for the entire range of multipoles is improved for both the $l3$ and the $l5$ model. Also, as expected, we see the biggest improvement at $l=3$ for the $l3$ model (red line), and a similar big improvement at $l=5$ for the $l5$ model (blue line). The reason for the huge improvement at $l=4$ is primarily the fact that for $n_{min}=2$ in Eq. \ref{par_est}, the sum in the numerator contains twice as many terms, as the sum in the denominator at $l=4$.  It may be possible that normalizations on other multipoles would produce stronger improvements in the parity parameter. Such an investigation will be the scope of a separate paper.\\
Finally, in order to investigate the impact of the results in Fig. \ref{fig:parity}, we have estimated the $p$-value to obtain a parity asymmetric power spectrum from pure random Gaussian and statistically isotropic CMB maps. The results for the $p$-value test are shown in Fig. \ref{parodd}. As one can see from this figure, de-correlation of the ILC 7 by the dipole modulated KBO can significantly increases the p-value by more than one order of magnitude for $l_{\mathrm{max}}\sim 20-25$. Though the actual changes in $g(l)$, as well as in the $p$-value, are still low, Fig. \ref{parodd} clearly shows the tendency, despite our relatively simple model for the KBO foreground (see discussion in section \ref{section:decor}).\\

\section{Conclusion}
In this paper, we have investigated a possible solution to the anomalous parity asymmetry in the Cosmic Microwave Background and the quadrupole-octupole alignment, related to the solar system foreground. The Kuiper Belt objects in the solar system plane can contribute to the microwave sky, and thus create a power contrast between even and odd multipoles. We have built a model of the KBO emissivity, based on the symmetry of the KBO in ecliptic coordinates. An essential part of the model is
the incorporation of the dipole modulation of the ILC 7 map and KBO foreground, which transform the symmetric KBO foreground into an asymmetric part. An important point of our analysis is that by maximizing the cross-correlations between that asymmetric component and the ILC 7 odd harmonics for $l=3$, and separately for $l=5$, we can fix the direction of dipole modulation, and significantly change the balance between even and odd multipoles in the ILC map.\\
To illustrate the possible changes to the morphology of the low multipole domain of the CMB power spectrum, we have applied the strongest de-correlations between the ILC 7 map and dipole modulated KBO foreground, in order to illuminate the clear improvement of the parity asymmetry and the quadrupole-octupole alignment.\\
We have developed a method for cleaning the ILC 7 $a_{l,m}$-values from the contribution of the dipole modulated KBO, thus retaining only the intrinsic CMB signal in the Ecliptic plane. It is possible to apply different weights to this filtering, allowing for various levels of parity symmetry for the low multipoles, depending on the normalization of the dipole modulation direction in the sky, and the corresponding amplitude of modulation. We have shown, that at the level where the parity is effectively restored, the quadrupole-octupole alignment would be significantly reduced down to the level of a chance correlation.\\
In conclusion, the removal of the KBO contribution, in the framework of the dipole modulation model, requires a statistical anisotropy of the CMB of non-cosmological origin. We believe, that the coming PLANCK data, with different systematics compared to the WMAP experiment, can put a light on the problem of low multipole anomalies in the CMB. 

\section*{Acknowledgments}
We are grateful to the anonymous referee, for very stimulating remarks and comments. We acknowledge the use of the Legacy Archive for Microwave Background Data Analysis (LAMBDA). Our data analysis made the use of HEALPix \cite{Healpix_primer} and GLESP \cite{Glesp}.  This work is supported in part by Danmarks Grundforskningsfond, which allowed the establishment of the Danish Discovery Center. This work is supported by FNU grant 272-06-0417, 272-07-0528 and 21-04-0355. CB acknowledges partial support by ASI through ASI/INAF Agreement I/072/09/0 for the Planck LFI Activity of Phase E2 and by MIUR through PRIN 2009.

\end{document}